\documentclass[a4paper,fleqn]{cas-dc}

\usepackage[numbers]{natbib}
\usepackage{url}
\usepackage{amsmath}
\usepackage{pifont}
\definecolor{ccr}{RGB}{10,110,150} 
\usepackage{hyperref}
\hypersetup{hypertex=true,
    colorlinks=pink,
    linkcolor=magenta,
    anchorcolor=blue,
    citecolor=blue,
    urlcolor=magenta
    }
\definecolor{light-gray}{gray}{0.8}    
\def\tsc#1{\csdef{#1}{\textsc{\lowercase{#1}}\xspace}}
\tsc{WGM}
\tsc{QE}
\tsc{EP}
\tsc{PMS}
\tsc{BEC}
\tsc{DE}

\begin{document}
\let\WriteBookmarks\relax
\let\printorcid\relax
\def\floatpagepagefraction{1}
\def\textpagefraction{.001}
\shorttitle{Leveraging social media news}
\shortauthors{Jie Zhou et~al.}

\title [mode = title]{MaDiNet: Mamba Diffusion Network for SAR Target Detection}                      


\author[1]{Jie Zhou}


\affiliation[1]{organization={College of Electronic Science and Technology},
                addressline={National University of Defense Technology}, 
                city={Changsha},
                postcode={410073}, 
                state={Hunan},
                country={China}}

\author[1]{Chao Xiao}


\author[1]{Bowen Peng}
\author[1]{Tianpeng Liu}
\author[1]{Zhen Liu}
\author[1]{Yongxiang Liu}
\author[1]{Li Liu}
\cormark[1]

\cortext[cor1]{Corresponding author}


\nonumnote{Email addresses: \url{zhoujie_@nudt.edu.cn} (J. Zhou), \url{xiaochao12@nudt.edu.cn} (C. Xiao), \url{pbow16@nudt.edu.cn} (B. Peng), 
\url{everliutianpeng@sina.com} (T. Liu), \url{zhen_liu@nudt.edu.cn} (Z. Liu), 
\url{lyx_bible@sina.com} (Y. Liu), \url{liuli_nudt@nudt.edu.cn} (L. Liu).
  }

\begin{abstract}
The fundamental challenge in SAR target detection lies in developing discriminative, efficient, and robust representations of target characteristics within intricate non-cooperative environments. However, accurate target detection is impeded by factors including the sparse distribution and discrete features of the targets, as well as complex background interference. In this study, we propose a \textbf{Ma}mba \textbf{Di}ffusion \textbf{Net}work (MaDiNet) for SAR target detection. Specifically, MaDiNet conceptualizes SAR target detection as the task of generating the position (center coordinates) and size (width and height) of the bounding boxes in the image space. Furthermore, we design a MambaSAR module to capture intricate spatial structural information of targets and enhance the capability of the model to differentiate between targets and complex backgrounds. The experimental results on extensive SAR target detection datasets achieve SOTA, proving the effectiveness of the proposed network. Code is available at \href{https://github.com/JoyeZLearning/MaDiNet}{https://github.com/JoyeZLearning/MaDiNet}.     

                 
\end{abstract}


\begin{highlights}
\item We introduce a diffusion-based detection method coupled with Mamba (MaDiNet), which conceptualizes the SAR target detection task as the generating target bounding box positions (center coordinates) and sizes (width and height) in the image space, significantly enhancing the inference accuracy of the detector. To the best of our knowledge, this is the first work to integrate diffusion models with Mamba for SAR target detection.

\item We design a MambaSAR module to capture rich and comprehensive spatial structure information about targets, emphasizing their positions within images and enhancing the detector to differentiate between targets and complex background interference.

\item We conduct a comprehensive analysis of the proposed MaDiNet against over 20 mainstream detection methods, including both anchor-based and anchor-free methods, across three representative datasets. Quantitative and qualitative experimental results demonstrate that MaDiNet achieves superior detection performance, surpassing all compared models. 

\end{highlights}


\maketitle

\section{Introduction}

Synthetic Aperture Radar (SAR) boasts a compelling capability for remote sensing imaging, offering high-resolution imagery that is largely impervious to variations in lighting and weather, thus establishing it as a vital tool for Earth Observation \cite{rizaev2022modeling,asiyabi2023synthetic,tuia2024artificial}. With the increasing amount of SAR image data, there is a pressing need to develop SAR image interpretation techniques, particularly SAR Automatic Target Recognition (ATR) methods \cite{chen2024reinforcement,zhou2024simulated}. SAR ATR is pivotal in a variety of applications, spanning both civilian domains such as advanced airport management and military operations, including reconnaissance and maritime monitoring \cite{li2024semi,zhou2024gaussian,zhang2024entropy}.  Consequently, it has been a vibrant field of research for several decades.

The objective of SAR target detection is to accurately and efficiently locate and recognize targets in complex scenes, and its core problem is developing discriminative, efficient, and robust representations of target characteristics. Recently, deep learning has brought significant progress for the problem of SAR target detection \cite{cui2024feature,huang2024physics,lv2024recognition}. In essence, detectors designed for SAR often draw upon methods established for object detection in natural images. However, the characteristics of SAR images are significantly distinct from those of natural images \cite{cheng2023towards,liu2020deep}. If detectors are used directly for SAR images without considering SAR characteristics, the effectiveness of the detection methods tends to be degraded. Currently, Convolutional Neural Network (CNN) based SAR target detection methods can be classified into anchor-based and anchor-free detectors. The former requires the placement of proposal boxes at predetermined locations to accurately locate targets, such as ShipDetector \cite{li2024lightweightsar}, MGANet \cite{ying2024mganet} and SFSCNet \cite{li2024unleashing}, while the latter detects targets through the organization of key points and includes keypoint based and center based methods, such as SARShipNet \cite{deng2022sarshipnet}, Improved FCOS \cite{yang2022improvedFcos}, and DCTC \cite{chen2024dctc}. In addition, vision transformers have also been introduced in SAR target detection \cite{lin2024dcea,li2024gl}. Despite these advancements, the following challenges persist and account for the inadequacies of the aforementioned  SAR target detection methods.

  \begin{figure}
	\centering
	\includegraphics[width=1\columnwidth]{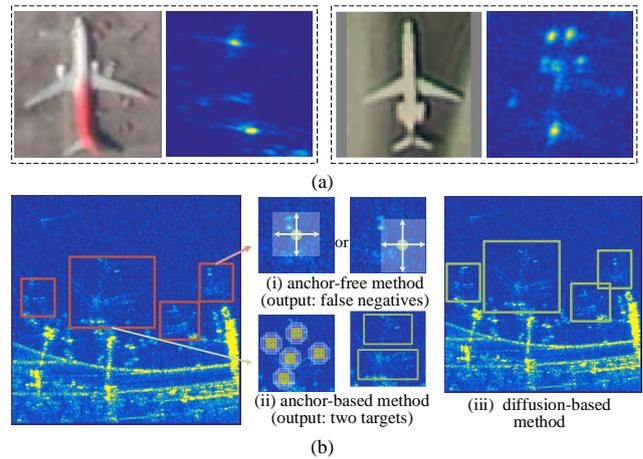}
	\caption{ (a) Examples of optical and SAR images for different aircrafts.  (b) (i) The traditional anchor-free method cannot accurately distinguish the central location and size of the discrete target and tends to generate false negatives.  (ii) The traditional anchor-based method detects a complete aircraft target as two separate aircrafts. (iii) The proposed diffusion-based detector is more flexible with discrete targets of various sizes. Red boxes are ground truth annotations and green boxes are prediction results. }
	\label{intro_discrete_challenge}
\end{figure}

\textbf{(i) Targets as a collection of discrete backscattering points.} Interested targets such as airplanes, ships, \emph{etc.} manifest themselves as a collection of discrete backscattering points in SAR images, resulting in a lack of fundamental appearance features like geometric shape, contour, and texture cues, as illustrated in Fig. \ref{intro_discrete_challenge} (a). These intrinsic characteristics raise challenges for target detection, especially for anchor-free methods that rely on keypoint localization. They can lead to the over-detection of targets and generate many false negatives, as depicted in Fig. \ref{intro_discrete_challenge} (b) (i). \emph{Therefore, how to develop novel methods to effectively explore the strong prior knowledge of SAR targets including the layout of the discrete backscattering points, the scattering characteristics, the salient target parts, and the surrounding environment of the targets is of great importance.}

\textbf{(ii) Diverse sizes and the sparse nature of SAR targets.} SAR images typically depict large scenes with sparse, diminutive targets, as shown in Table \ref{datasets_comparison}. Traditional anchor-based detection methods employ fixed, uniformly distributed proposal boxes across the feature map, leading to increased false positives and under-detection of irregularly shaped targets, as illustrated in Fig. \ref{intro_discrete_challenge} (b)(ii). They also raise computational complexity and resource usage due to redundant boxes. \emph{Thus, exploring more flexible approaches to anchor box generation is essential for accurate target detection.}

\begin{table}
	\centering
  \tabcolsep=1pt
	\begin{tabular}{ccccc}
 \toprule
		\multicolumn{2}{c}{Datasets}      & Img size  & BBox/Img & Ins/Img \\
  \midrule
	Natural & MS COCO 2014  & 640*480  & 3.9\%   & 7.3      \\
  \hline
\multirow{4}{*}{SAR} & SAR-ShipDataset & 256*256           & 1.8\%    & 3.77     \\
& \multirow{2}{*}{SAR-AIRcraft1.0} & 800*800$\sim$ & \multirow{2}{*}{1.0\%}     & \multirow{2}{*}{1.20}      \\
        &  & 1500*1500 &    &       \\
     & HRSID   & 800*800    & 0.12\%   &3.02  \\  
                    \bottomrule
	\end{tabular}
	\caption{Comparison between natural and SAR detection datasets (MS COCO 2014 \cite{lin2014microsoft}, SAR-ShipDataset \cite{wang2019sarshipdataset}, SAR-AIRcraft1.0 dataset \cite{zhirui2023sarcraft}, HRSID \cite{wei2020hrsid}). BBox/Img denotes the mean proportion of the area of the target bounding box relative to the total area of the image. Ins/Img represents the mean number of target instances per image. (BBox, Ins, and Img are abbreviations for bounding boxes, instances, and images, respectively.)}
	\label{datasets_comparison}
\end{table}

\begin{figure}[t]
	\centering
	\includegraphics[width=1\columnwidth]
{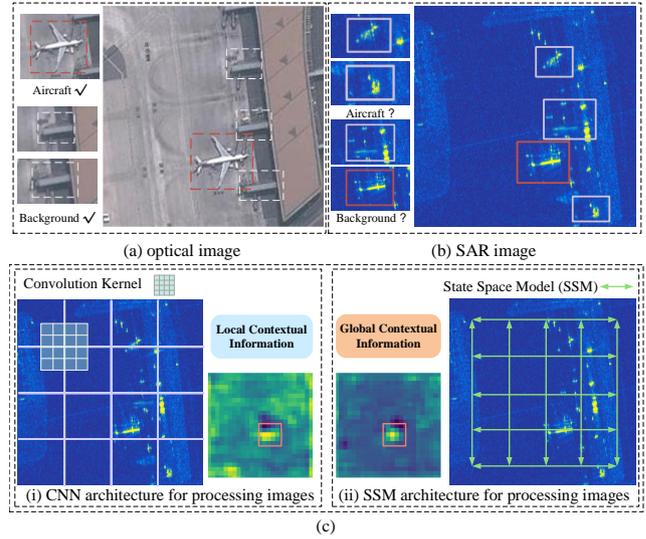}
	\caption{  (a) optical image. Detectors can easily detect the target based on its appearance. (b) SAR image.  Detectors with a limited receptive field may easily lead to incorrect results. (c) Feature maps are generated by MaDiNet without SSM (i) and with SSM (ii).  The wide range of contextual information around the target conduces to suppress the background interference and enhance the target representation. Red boxes are ground truth.}
	\label{mambaAndFeature}
\end{figure}

 \textbf{(iii) Complex scenes and background interference.} As illustrated in Fig. \ref{mambaAndFeature} (a) (b), the scattering characteristics of surrounding structures and metal objects are highly similar to the target, making feature-based discrimination inadequate for separating the aircraft from the background clutter. \emph{Therefore, harnessing contextual information to reduce background interference and sharpen target feature representation is important for accurate SAR target detection.}

\begin{figure}[t]
	\centering
	\includegraphics[width=1\columnwidth]
{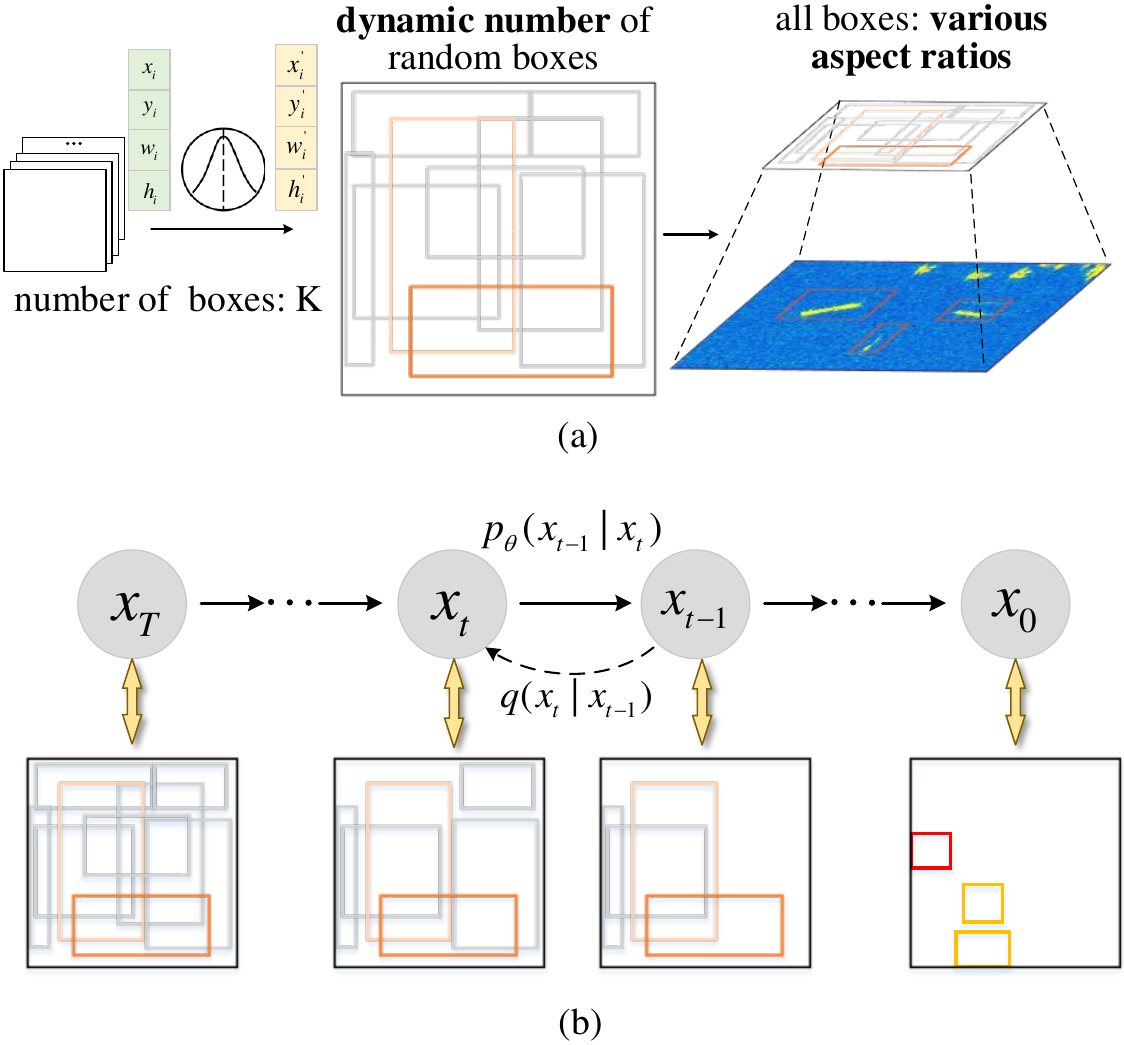}
	\caption{ (a) random boxes (Detection from random boxes without priors, and x,y,w,h represent the positions (center coordinates) and sizes (widths and heights) of bounding boxes in the image.) (b) Simplified schematic diagram of our proposed MaDiNet detection process. We only apply noise addition and denoising to the target detection boxes, without denoising the SAR images.
}
	\label{randomBoxDiff}
\end{figure}

\begin{figure}[t]
	\centering
	\includegraphics[width=0.9\columnwidth]
{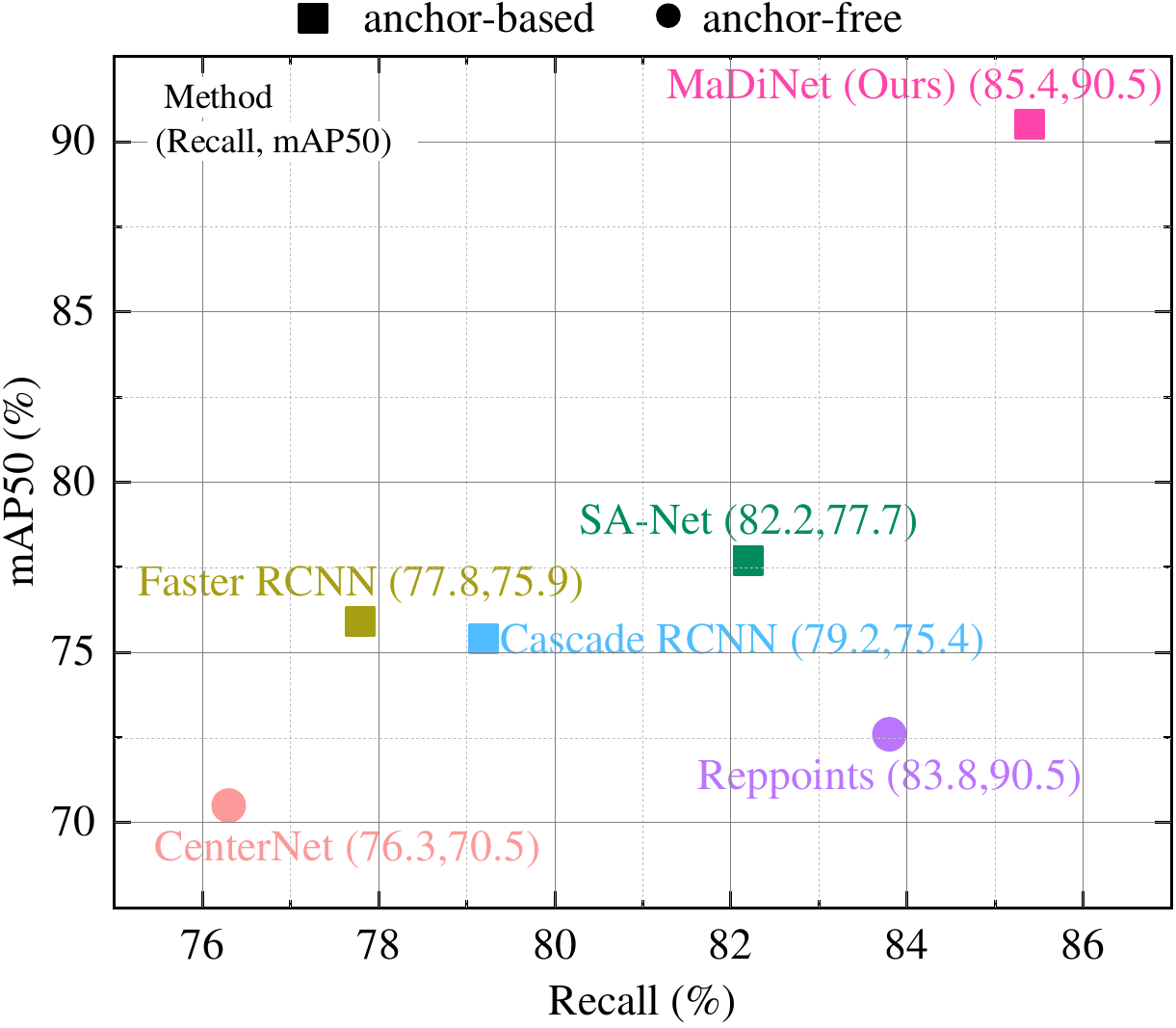}
	\caption{ Comparison of the detection performance (Recall and mean average precision) of different methods on the SAR-AIRcraft1.0 dataset \cite{zhirui2023sarcraft}. Our proposed method is highly effective and covers a wider range of targets compared with other classical methods, including anchor-based (Faster RCNN \cite{ren2016fasterrcnn}, Cascade RCNN \cite{cai2019cascadercnn}, and SA-Net \cite{{zhirui2023sarcraft}}), and anchor-free (CenterNet \cite{duan2019centernet} and Reppoints \cite{yang2019reppoints}) methods.}
	\label{diffMap}
\end{figure}

Inspired by DiffusionDet \cite{chen2023diffusiondet}, this paper conceptualizes SAR target detection as a denoising diffusion process
from noisy boxes to target boxes, as illustrated in Fig. \ref{randomBoxDiff}. The detector operates independently of prior knowledge or manual design of anchor boxes. It directly samples sizes and positions of boxes from a Gaussian distribution, effectively handling scale variations of the target and preventing the separation of critical scattering points. Furthermore, to overcome the limited receptive fields of CNN-based methods, we design a MambaSAR module based on the State Space Model (SSM) with robust long-range dependency perception capabilities and construct a correlation between key scattering points of targets and their rich contextual information. As depicted in Fig. \ref{mambaAndFeature} (c), this interactive correlation facilitates the dynamic capture of both global contextual information and local features amidst complex backgrounds. The design effectively suppresses background interference and acquires comprehensive semantic structure information about targets, thereby enhancing overall detection performance. Fig. \ref{diffMap} presents the detection results of different detection methods, demonstrating the advantage of our proposed diffusion-based MaDiNet over existing methods with high mean precision and recall rate.

Our initial exploration of diffusion-based detectors was published as DiffDet4SAR  \cite{zhou2024diffdet4sar}. In this extended version, we improve the work based on the following key aspects: (i) The original DiffDet4SAR only utilized the classic ResNet50 as its backbone network. Given the background interference present in SAR images, we expand DiffDet4SAR by integrating the SSM structure into the network, enabling our proposed MaDiNet to accurately detect targets within more complex scenarios. (ii) We conduct a systematic investigation of detection results using various backbone networks, including ConvNext, Transformer, and MambaVision. Experimental results indicate that our designed backbone network incorporating an SSM can significantly enhance target feature representation and improve detection performance. (iii) While the original DiffDet4SAR is tested solely on a high-resolution aircraft target dataset, this version of MaDiNet undergoes extensive evaluation across multiple datasets with varying resolutions and sizes that include diverse targets and achieves better accuracy than existing detectors. The primary contributions of this work can be summarized as follows:

(1) We introduce a diffusion-based detection method coupled with Mamba, named MaDiNet, which conceptualizes the SAR target detection task as the generating target bounding box positions (center coordinates) and sizes (width and height) in the image space, significantly enhancing the inference accuracy of the detector. To the best of our knowledge, this is the first work to integrate diffusion models with Mamba for SAR target detection.
    
(2) We design a MambaSAR module to capture rich and comprehensive spatial structure information about targets, emphasizing their positions within images and enhancing the detector to differentiate targets and complex background interference.
    
(3) We conduct a comprehensive analysis of the proposed MaDiNet against over 20 mainstream detection methods, including both anchor-based and anchor-free methods, across three representative datasets. Quantitative and qualitative experimental results demonstrate that MaDiNet achieves superior detection performance, surpassing all compared models.

\section{Related works}\label{relatedwork}

\subsection{SAR Target Detection}
The origins of radar target detection algorithms can be found in \cite{finn1966adaptiveDetection, hm1968cacfar}, which investigated and tackled the problem of automatically identifying point targets in an extended clutter environment that is not smooth. Later on, Lincoln Laboratory \cite{novak1993SARATR} presented a three-level process that separated the tasks of detection and recognition. As more and more cutting-edge methods become available, the task of target detection has progressed from localization to fine-grained recognition. Deep learning-based SAR target detection methods have become the most popular among these developments due to their strong feature learning capabilities and excellent detection performance. Based on the use of anchors, these methods can be divided into anchor-based and anchor-free methods. Because anchor-free methods have difficulties in adapting to the discrete features of SAR targets, their performance still needs to be improved \cite{deng2022sarshipnet,chen2024dctc}. Therefore, here We concentrate on anchor-based SAR target detection methods.

The core problem of SAR target detection is developing discriminative, efficient, and robust representations of target characteristics. GEFCSI-Net \cite{yang2023GEFCSI} created a multi-scale adaptive feature pyramid network to achieve dense information interaction and cross-scale feature fusion between various feature maps. MLSDNet \cite{chang2023mlsdnet} aggregated the position and contour features of targets by using depth-wise separable convolution to obtain scattering information of multi-scale targets. However, as the network gets deeper, these methods might lose effective features. To help the model better capture spatial features and location information in images, OE-YOLO \cite{wu2024object} added a coordinate attention mechanism to the YOLOv7 \cite{yolov7} backbone. SFS-CNet \cite{li2024unleashing} mapped the input feature into spatial and frequency components. The former uses a fractional-order Gabor transformer to capture rich frequency variations and texture features, while the latter uses dynamic receptive field adjustment to perceive contextual information of different objects. Furthermore, SA-Net \cite{zhirui2023sarcraft} sought to increase aircraft target localization accuracy through the use of scatter key points for positioning. Nevertheless, these methods usually distribute anchor boxes uniformly in feature maps and predefine their shape and location based on natural image priors. The performance of detectors is constrained by their tendency to overlook the sparse characteristics of SAR images, resulting in missed targets with irregular shapes or disconnected scatter points.

\subsection{Diffusion Models}

Diffusion models, as a form of probabilistic generative models, have demonstrated significant success in the realm of computer vision \cite{kolbeinsson2024multi, shen2024diffclip}, including image generation, image restoration, and modality translation. Drawing inspiration from non-equilibrium statistical physics, Sohl-Dickstein et al. \cite{sohl2015deep} initially proposed the diffusion model by systematically and iteratively deconstructing the structure of data distribution through a forward diffusion process. Serving as the foundation of diffusion models, DDPM \cite{ho2020denoising} has made substantial contributions to this field by employing variational inference for modeling, including GLIDE \cite{nichol2021glide}, Stable Diffusion \cite{rombach2022high}, and SDXL \cite{podell2023sdxl} models. In image generation, Denisa Qosja et al. \cite{qosja2024sar} were the first to explore the application of DDPM for generating SAR images and discussed various design choices and parameters to adapt it to the SAR image characteristics. In terms of image restoration, SAR-DDPM \cite{perera2023sar} was the pioneering work in utilizing DDPM for despeckling SAR images through iterative prediction of added noise using a noise predictor conditioned on speckled images. Hu et al. \cite{hu2024sar} developed the region-denoising diffusion probability model R-DDPM for SAR images, which involves segmenting the image into specific-sized regions to achieve high-quality despeckling without edge artifacts. In the modal translation, Guo et al. \cite{guo2024learning} considered the color correlation and direct mapping between SAR and optical image domains, proposing a color memory diffusion model (CM-Diffusion) for SAR-to-optical image translation (S2OIT)  to address color memory in different scenes. Gou et al. \cite{gou2024interpretable}, guided by the interpretable latent space of diffusion models, constructed an optical-SAR image translation and matching framework through a dynamic conditional diffusion model (DCDM) to achieve interpretable and robust cross-modal image matching between optical and SAR images. In terms of perceptual task, inspired by DiffusionDet \cite{chen2023diffusiondet}, Zhou et al. \cite{zhou2024diffdet4sar} first applied the diffusion model into SAR image target detection, conceptualizing the target detection task as a generation task from noisy boxes to ground truth boxes. It is worth noting that the potential of diffusion models in SAR image interpretation remains an area that has not been thoroughly explored.

\subsection{Mamba}

Mamba \cite{gu2023mamba} is a representative selective structured State Space Model (SSM) \cite{kalman1960newlinear, gu2021efficientlyssm} that has achieved promising performance on long sequence modeling tasks with linear complexity \cite{hatamizadeh2024mambavision, wang2024mambayolo, zhu2024visionmamba}. In the field of remote sensing, Pan-Mamba \cite{he2024pan} first attempted to introduce the Mamba model into the pan-sharpening domain. RSMamba \cite{chen2024rsmamba} designed a multipath visual state space (VSS) block for large-scale image interpretation. Ma et al. proposed RS3Mamba \cite{ma2024rs3mamba} for image semantic segmentation by using VSS blocks to construct an auxiliary branch. These methods highlight the significant potential of Mamba, and its particular application in SAR target detection tasks remains unexplored.

\section{Methods}\label{methods}

\begin{figure*}
	\centering
        \includegraphics[width=2\columnwidth]{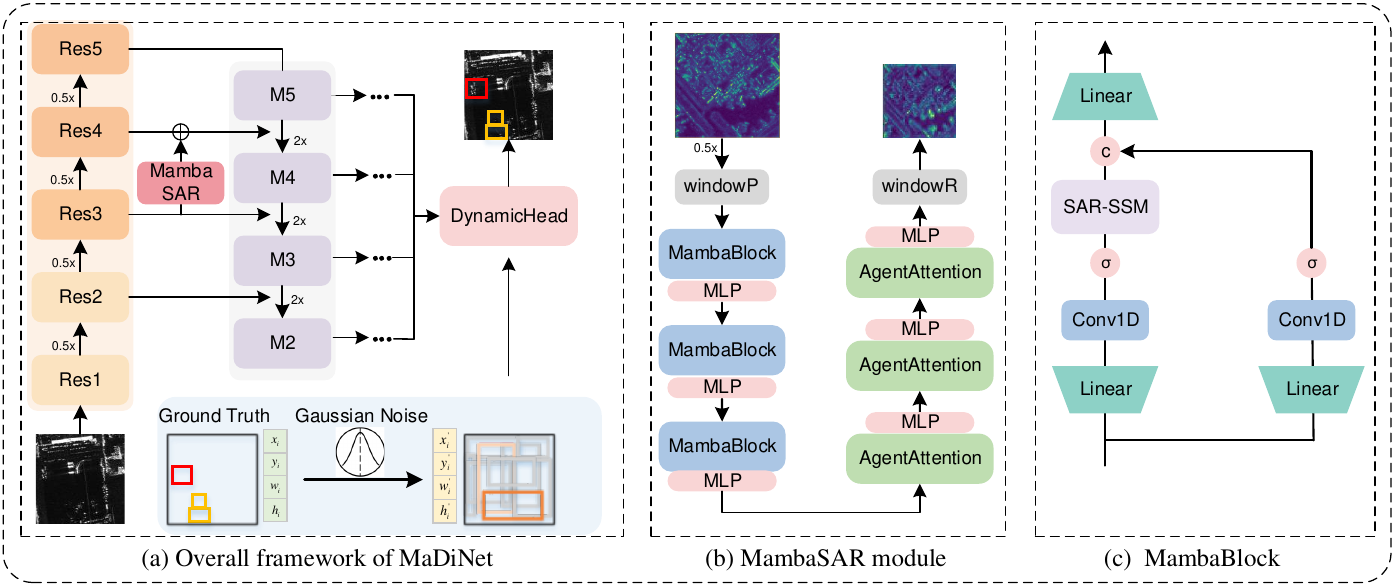}
	\caption{(a) MaDiNet framework for SAR target detection, featuring a comprehensive diffusion-based noise-to-box paradigm with a backbone network incorporating MambaSAR module. (M2-M4 denotes the intermediate feature layers within the feature pyramid network.) (b) The MambaSAR module is a hybrid architecture consisting of mamba blocks, agent attention blocks, and MLP blocks. The module first employs an adaptive window-scratching mechanism to tailor SAR images of multiple sizes. (c) The architecture of MambaBlock in the MambaSAR module. In addition to replacing the causal Conv layer with its regular counterparts, we create a symmetric path without SAR-SSM as a token mixer to enhance the modeling of global features.}      
	\label{framework_mdiffdet4sar}
\end{figure*}

\subsection{The overall framework of MaDiNet}

In this paper, we describe SAR target detection as the task of generating the position (center coordinates) and size (width and height) of the bounding boxes in the image space. The proposed method does not require prior candidate boxes, alleviating the inaccurate positioning and false negatives caused by various sizes of targets.

As shown in Fig. \ref{randomBoxDiff} (b), Gaussian noise is added to the ground truth box during the training phase to obtain a noisy box with variance scheduling control. During the inference phase, MaDiNet generates bounding boxes by reversing the learned diffusion process, which adjusts the noise prior distribution to the distribution of the learned bounding boxes.

The learning objective for SAR target detection is the input-target pairs $(x, b, c)$, where $x$ is the input image, and $b$ and $c$ are the bounding box and class label in SAR image $x$, respectively. Specifically, we represent the $i$-th box in the set as $b^{i}$ = $(c_x^{i},c_y^{i},w^{i},h^{i})$, where $(c_x^{i},c_y^{i})$ are the center coordinates of the bounding box and $(w^{i},h^{i})$ are its width and height. The forward noise process can be defined as:
\begin{equation}
\begin{aligned}
q\left(z_t \mid z_{0}\right)=\mathcal{N}\left(z_t ; \sqrt{\bar{\alpha_t}} z_{0},\left(1-\bar{\alpha_t}\right) \mathbf{I}\right),
\end{aligned}
\end{equation}
which transforms data sample $z_{0}$ into a latent noisy sample $z_{t}$ for $t \in [0,1,...,T]$ by adding noise to $z_{0}$, with $\bar{\alpha_{t}}= \alpha_{1}\cdot \alpha_{2}...\cdot \alpha_{t}$. 

In our setting, the data sample is a set of bounding boxes $z_0 = b$, where $b\in R^{N\times 4}$ represents a collection of $N$ bounding boxes (the four dimensions represent the center coordinates, width, and height of the boxes). The neural network $f_{\theta } (z_t,t,x)$ is trained to predict $z_0$ and generate the corresponding class label $c$ from the noisy box $z_t$ by minimizing the training objective in terms of $L_{2}$ loss:
\begin{equation}
\begin{aligned}
\mathcal{L}_{\text {train }}=\frac{1}{2}\left\|f_\theta\left(\boldsymbol{z}_t, t\right)-\boldsymbol{z}_0\right\|^2.
\end{aligned}
\end{equation}

In the backbone depicted in Fig. \ref{framework_mdiffdet4sar} (a), the core component incorporates the MambaSAR module to capture richer contextual information. This module first downsamples features, then uses agent attention to highlight local key features and acquire global features via MambaBlock. These features are then combined with the original layer 4-th features to create new high-dimensional image representations. The neck of the network is designed with Feature Pyramid Networks (FPN). Furthermore, a fixed convolutional kernal may not be easily adapted to discontinuous and irregular targets due to the wide range of target scales in SAR images. To address this, we employ flexible deformable convolution in the middle and high-dimensional feature layers. Unlike standard convolution, this deformable convolution enables the model to learn overall features that better fit the actual shapes of targets by allowing adjustments to be made to the sampling positions of the convolution kernels. This improves the effectiveness of feature extraction by capturing more meaningful information.

\subsection{MambaSAR Module}
Inspired by MambaVision \cite{hatamizadeh2024mambavision}, we propose a hybrid architecture that integrates Mamba blocks and attention blocks for SAR target detection task, as illustrated in Fig. \ref{framework_mdiffdet4sar} (b). This design incorporates an adaptive window-scratching mechanism, allowing it to better accommodate SAR images of various sizes. Moreover, utilizing multiple self-attention blocks in the final stages significantly enhances the model's ability to capture global context and long-range spatial dependencies. 

Assuming an input $\boldsymbol{x}\in R^{T\times C}$ with sequence length $T$ with embedding dimension C, we design 6 layers in the MambaSAR module, and the output of layer $n$ can be computed as 

\setlength{\abovedisplayskip}{0.5pt}
\setlength{\belowdisplayskip}{0.1pt}
\begin{equation}
\begin{aligned}
& \hat{\boldsymbol{x}}^{n} = MambaBlock(Norm(\boldsymbol{x}^{n-1}))+\boldsymbol{x}^{n-1},\qquad n=0,1,2,\\
& \hat{\boldsymbol{x}}^{m} = AttentionBlock(Norm(\boldsymbol{x}^{m-1}))+\boldsymbol{x}^{m-1},  \quad n=3,4,5,\\
& \boldsymbol{x}^{i} = MLP(Norm(\hat{\boldsymbol{x}}^{i} ))+\hat{\boldsymbol{x}}^{i},\qquad \quad \ \, \qquad \qquad  i=0,1...5,
\end{aligned}
\end{equation}
where Norm, MambaBlock, and AttnetionBlock denote the choices of layer normalization and token blocks, respectively. Without loss of generality, Layer Normalization is used for Norm. Given 6 layers in the MambaSAR module, the first 3 layers
employ MambaBlocks while the remaining 3 layers employ Agent-attention blocks. We describe the details of each block in the following.

\subsubsection{MambaBlock}
Given that SAR imaging fundamentally represents a mapping from the target scattering property space to the target image space, we have redesigned the original MambaBlock, as illustrated in Fig. \ref{framework_mdiffdet4sar} (c). 

Initially, we suggest swapping out causal convolution for regular convolution because the former limits influence to one direction. Second, given the intricate structures of targets in SAR images and the extreme sensitivity of the scattering center intensity to the azimuth angle, we present a SAR-SSM module that utilizes a four-directional scanning mechanism. This method collects rich spatial information from multiple angles and allows for global modeling of features across multiple orientations. Furthermore, targets in SAR images are usually discontinuous, consisting of several distinct bright spots with asymmetric scattering centers. It is crucial to integrate these elements for effective recognition. Therefore, we incorporate a symmetric branch without SAR-SSM that consists of additional convolutions and SiLU activation functions to compensate for any content loss due to sequential constraints inherent in SAR-SSM. Ultimately, we combine the outputs from both branches and apply a linear layer to them. By utilizing the benefits of each branch, this combination makes sure that our final feature representation preserves both global and local information.

Given an input $\boldsymbol{x}_{in}$, the output of MambaBlock $\boldsymbol{x}_{out}$ is computed according to   

\begin{figure*}
	\centering
	\includegraphics[width=2\columnwidth]
{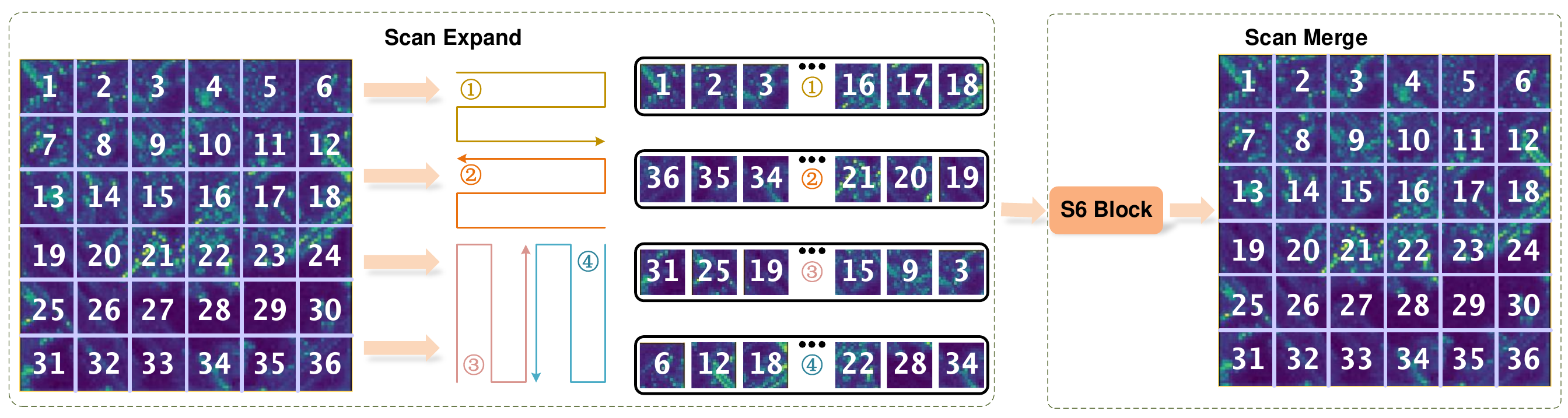}
	\caption{Illustration of SAR-SSM operation. The scan expansion process in SAR-SSM is divided into four branches, each scanning the feature map block by block along distinct paths to generate four separate sequences. The subsequent scan merge operation utilizes these sequences as inputs to the S6 block, where they are merged from different directions, allowing for the extraction of global features.}
	\label{mambascan}
\end{figure*}

\setlength{\abovedisplayskip}{0.5pt}
\setlength{\belowdisplayskip}{0.1pt}
\begin{equation}
\begin{aligned}
& \boldsymbol{x}_{1} = SARScan(\sigma(Conv(Linear(C,\dfrac{C}{2})(\boldsymbol{x}_{in})))),\\
& \boldsymbol{x}_{2} = \sigma(Conv(Linear(C,\dfrac{C}{2})(\boldsymbol{x}_{in}))),\\
& \boldsymbol{x}_{out} = Linear(\dfrac{C}{2},C)(Concat(\boldsymbol{x}_{1},\boldsymbol{x}_{2}),
\end{aligned}
\end{equation}
where linear ($C_{in}$,$C_{out}$)($\cdot$) denotes a linear layer with $C_{in}$ and $C_{out}$ as input and output embedding dimensions. SARScan is our proposed Multi-directional selective scan operation as shown in Fig. \ref{mambascan}, which will be introduced in the next part. $\sigma$ is the activation function for which the Sigmoid Linear Unit (SiLU) \cite{elfwing2018silu} is used. In addition, Conv and Concat represent 1D convolution and concatenation operations.

\subsubsection{SAR-SSM}
Due to the complex structure of targets, different parts exhibit varying scattering mechanisms, including cavity scattering and edge-winding. During the SAR imaging process, these scattering characteristics change to different extents based on conditions such as incidence angle and azimuth angle, resulting in variable scattering characteristics that make target classification difficult. To address this challenge, we design the SAR-SSM scanning merging mechanism to capture enhanced features of targets from multiple directions and angles. The proposed SAR-SSM consists of three main steps: scan expansion, S6 chunking \cite{gu2023mamba}, and scan merging, with its overall flow illustrated in Fig. \ref{mambascan}. 

The scan expansion operation unfolds the input image into a series of sub-images, each representing a specific direction. From a diagonal perspective, this process moves in four symmetric directions: left-to-right, top-down, bottom-up, and right-to-left. This layout not only ensures comprehensive coverage of all regions within the input image but also establishes a rich multidimensional information base for subsequent feature extraction through systematic orientation transformation. Consequently, the orientation sensitivity in SAR images can be effectively reduced by using our suggested four-direction scan operation.

\subsubsection{Agent Attention}
\begin{figure}
	\centering
	\includegraphics[width=1\columnwidth]
{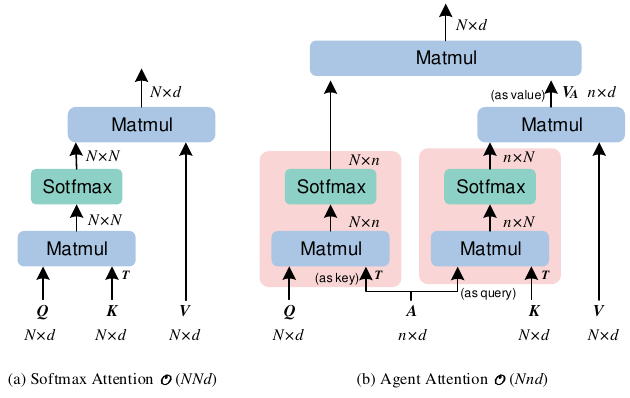}
	\caption{Difference between softmax attention and agent attention. (a) Softmax attention computes similarities between all query-key pairs, resulting in quadratic complexity concerning the number of tokens. This can lead to inefficiencies, especially in scenarios involving long sequences. (b) Agent attention utilizes a small group of agent tokens to aggregate and broadcast global information. This approach elegantly integrates the benefits of both softmax and linear attention, achieving high expressiveness while maintaining low computational complexity.}  
	\label{agentattention}
\end{figure}

Mamba provides linear complexity by compressing the filtered global context into a hidden state. However, this compression invariably results in the loss of fine-grained local dependency information between tokens, which has an impact on the accuracy of SAR target detection. To solve this problem, we design an attention structure inside the MambaBlock backend that concentrates on various areas of the image to more accurately represent global information. As a result, the model is better equipped to discern between targets and interferences from complex backgrounds and can learn representations with greater fineness. The structure of agent attention is illustrated in Fig. \ref{agentattention}.

Traditional softmax and linear attention methods either exhibit high computational complexity or lack sufficient representation ability. Historically, these two paradigms have been approached as separate approaches in previous research with the goal of either optimizing linear attention performance or lowering the computational cost of softmax attention. Here, we incorporate agent attention \cite{han2023agentattention} into our MambaBlock, which exploits redundancy between attention weights to achieve
both high model expressiveness and low computation complexity. Agent attention elegantly integrates softmax and linear mechanisms, benefiting from their high expressiveness alongside linear complexity. 

\setlength{\abovedisplayskip}{0.5pt}
\setlength{\belowdisplayskip}{0.1pt}
\begin{equation}
\begin{aligned}
&  Atten(Q,K,V)=Softmax(\frac{QK^{T}}{\sqrt d_{h}})V,  
\end{aligned}
\end{equation}
where $Q$,$K$,$V$ denote query, key and value, respectively. And $d_{h}$ is the number of attention heads.  

\setlength{\abovedisplayskip}{0.5pt}
\setlength{\belowdisplayskip}{0.1pt}
\begin{equation}
\begin{aligned}
& AgentAtten(Q,A,K,V)=Atten(Q,A,Atten(A,K,V)),  
\end{aligned}
\end{equation}
where $A\in \mathbb{R}^{n\times C}$ is the newly defined agent tokens. Specifically, agent tokens $A$ are treated as queries, and attention between $A$, $K$, and $V$ are calculated to aggregate agent features $V_{A}$ from all values. Subsequently, $A$ are utilized as keys and $V_{A}$ as values in the second attention calculation with the query matrix $Q$, broadcasting the global information from agent features to every query token and obtaining the final output. In this way, the computation of pairwise similarities between $Q$ and $K$ can be avoided while information exchange between each query-key pair through agent tokens is preserved.   

\section{Experimental settings}\label{experimentsetting}

\subsection{Datasets}
To demonstrate the effectiveness of the proposed MaDiNet, three open-source datasets of different categories, sizes, and scales have been selected for performance evaluation, including  SSDD \cite{li2017shipssdd, zhang2021sarssdd}, SAR-ShipDatatset \cite{wang2019sarshipdataset} and SAR-AIRcraft1.0 dataset \cite{zhirui2023sarcraft}, as shown in Table \ref{DataSets}. The details of these datasets are illustrated as follows.

\textbf{1) SSDD.} The SSDD\footnote{download link: \url{https://github.com/TianwenZhang0825/Official-SSDD
}} is the first publicly available SAR detection dataset, which was first introduced by the Naval Aeronautical Engineering Academy in 2017 at the BIGSARDATA conference \cite{li2017shipssdd}.  The SAR image samples in the SSDD originate from different sensors and exhibit various resolutions. The dataset includes images with different polarizations under various sea conditions, and depicting different ship scenarios, encompassing both coastal and offshore areas, as well as ships of varying sizes. The SSDD \cite{zhang2021sarssdd} used in this paper is the officially revised version re-annotated and standardized by the Chinese Academy of Sciences in 2021, which contains 1160 images and 2456 ship instances.

\begin{table*}[!ht]
\setlength{\tabcolsep}{2.5pt}
    \centering
     \caption{Details of the datasets used in the experiments. SSDD is the first publicly available SAR detection dataset, SAR-AIRcraft1.0 is the first aircraft detection dataset containing fine-grained recognition, and SAR-ShipDataset is the largest publicly available SAR ship detection dataset.
     }\label{DataSets}
     \scalebox{0.9}{
    \begin{tabular}{ccccccccccc}
    \hline
        Datesets & Target & Classes & Resolution(m) & Band & Polarization & Satellites & Size & Images & Instances & Ins./Img. \\ \hline
        \multirow{2}{*}{SSDD \cite{zhang2021sarssdd}} & \multirow{2}{*}{ship} & \multirow{2}{*}{1} & \multirow{2}{*}{1$\sim$15} & \multirow{2}{*}{C/X} & HH,VV, & Sentinel-1 & 190x214$\sim$& \multirow{2}{*}{1160} & \multirow{2}{*}{2456} & \multirow{2}{*}{2.12} \\ 
        ~ & ~ & ~ & ~ & ~ & VH,HV & RadarSat-2,TerraSAR-X & 526x668 & ~ & ~ & ~ \\ 
        
        \multirow{2}{*}{SAR-AIRcraft1.0 \cite{zhirui2023sarcraft}} & \multirow{2}{*}{aircrafts} & \multirow{2}{*}{7} & \multirow{2}{*}{1} & \multirow{2}{*}{C} & \multirow{2}{*}{Uni-Polar} &  \multirow{2}{*}{Gaofen-3} & 800x800$\sim$ & \multirow{2}{*}{4368} & \multirow{2}{*}{16463} & \multirow{2}{*}{3.77} \\ 
         ~ & ~ & ~ & ~ & ~ & ~ & ~ & 1500x1500 & ~ & ~ & ~ \\ 
        \multirow{2}{*}{SAR-ShipDataset \cite{wang2019sarshipdataset}}  & \multirow{2}{*}{ship} & \multirow{2}{*}{1} & \multirow{2}{*}{3$\sim$25} & \multirow{2}{*}{C} & HH,VV, &  Gaofen-3,  & \multirow{2}{*}{256x256} & \multirow{2}{*}{39729} & \multirow{2}{*}{47416} & \multirow{2}{*}{1.20} \\ 
         ~ & ~ & ~ & ~ & ~& VH,HV& Sentinel-1 & ~  & ~ & ~ & ~ \\ \hline
    \end{tabular}
}
\end{table*}

\begin{table}
\setlength{\tabcolsep}{2.5pt}
    \centering
    \caption{Comparisons of detection performance on SSDD \cite{zhang2021sarssdd}. Results are computed by precision (P (\%)), recall (R (\%)), mAP$_{50}$ (\%) and F1-score (\%). The \textbf{best} and \underline{second best} results are shown in \textbf{boldface} and \underline{underline} (confidence=0.5). $\diamond$ represents anchor-free methods. }\label{tabel_ssdd_results}

    \begin{tabular}{cccccc}
  
\toprule
  \multirow{2}{*}{Methods} & Open- &  \multirow{2}{*}{R} & \multirow{2}{*}{P} & \multirow{2}{*}{AP$_{50}$} &\multirow{2}{*}{F1-score} \\
           ~ &source&~&~~&~\\ \midrule
          $\diamond$SAR-ShipNet\cite{deng2022sarshipnet} &\ding{56}& 76.3 & 95.1 & 89.1 & 85.0 \\ 
             $\diamond$DCTC \cite{chen2024dctc} & \ding{56}&- & - & 93.9 & - \\
         
              GEFCSI-Net \cite{yang2023GEFCSI} &\ding{56}& 96.9 & 93.1 & 97.1 & 95.0 \\
         SARNas \cite{du2023sarnas} & \ding{56}& 94.8 & 97.4 & 98.5 & 96.1 \\ 
         MGA-Net \cite{ying2024mganet} &\ding{56}& \textbf{99.0} & 94.7 & \underline{98.6} & 96.8 \\ 
    \hline
         Faster RCNN \cite{ren2016fasterrcnn} & \ding{52}&90.4 & 87.1 & 89.7 & 88.7 \\ 
        Cascade RCNN \cite{cai2019cascadercnn} &\ding{52}& 90.8 & 84.1 & 90.5 & 92.4 \\ 
      
         RetinaNet\cite{lin2017focal} &\ding{52}& 93.1 & 94.0 & 95.0 & 93.6 \\ 
            $\diamond$CenterNet \cite{duan2019centernet} & \ding{52}&93.6 & 94.3 & 95.1 & 93.9 \\ 
         $\diamond$FCOS \cite{tian2019fcos} &\ding{52}& 93.9 & 94.6 & 95.3 & 94.2 \\ 
          ConsistencyDet \cite{jiang2024consistencydet} &\ding{52}& 94.0&94.5&95.6& 94.2\\
         YOLOv3 \cite{redmon2018yolov3} & \ding{52}&94.8 & 95.5 & 96.2 & 95.2 \\ 
         YOLOX \cite{ge2021yolox} & \ding{52}&93.8 & 95.0 & 96.2 & 94.4 \\
 DiffusionDet \cite{chen2023diffusiondet} &\ding{52}&93.9&95.1&96.3&94.5\\
         YOLOv5n \cite{yolov5} & \ding{52}&95.8 & 97.0 & 98.0 & 96.4 \\ 
             YOLOv8n \cite{yolov8} & \ding{52}&96.4 & \underline{97.3} & 98.1 & 96.9 \\ 
        YOLOv5s \cite{yolov5} & \ding{52}&96.4 & 97.2 & 98.3 & \underline{97.0} \\

        DiffDet4SAR \cite{zhou2024diffdet4sar} & \ding{52}&95.2& 95.3&96.9&95.2\\

         MaDiNet(Ours) & \ding{52}& \underline{97.0}& \textbf{98.0}&\textbf{99.0}& \textbf{97.5}
       
       \\

 \bottomrule
    \end{tabular}
\end{table}

\begin{table}
    \setlength{\tabcolsep}{3.5pt}
    \centering
    \caption{Comparisons of detection performance on SAR-ShipDataset \cite{wang2019sarshipdataset}. Results are computed by precision (P (\%)), recall (R (\%)), mAP$_{50}$ (\%) and F1-score (\%). The \textbf{best} and \underline{second best} results are shown in \textbf{boldface} and \underline{underline} (confidence=0.5). $
\diamond$ represents anchor-free methods. }\label{table_shipdata_results}
    \begin{tabular}{cccccc}
    
\toprule
          \multirow{2}{*}{Methods} & Open- &  \multirow{2}{*}{R} & \multirow{2}{*}{P} & \multirow{2}{*}{AP$_{50}$} &\multirow{2}{*}{F1-score}\\
           ~ &source&~&~~&~&~\\ \midrule
         OE-YOLO \cite{wu2024object} & \ding{56}& 81.5 & 85.4 & 86.0 &83.4\\ 
            $\diamond$ SAR-ShipNet \cite{deng2022sarshipnet} &\ding{56}&71.3 &94.9&90.2&81 \\
         ShipDetector \cite{li2024lightweightsar} & \ding{56}& - & - & 91.2&- \\

       $\diamond$ Improved FCOS \cite{yang2022improvedFcos} & \ding{56}& 95.2 & 89.8 & 94.1&92.4 \\ 

        GEFCSI-Net \cite{yang2023GEFCSI}& \ding{56} & 93.0 & 90.2 & 94.3&91.6 \\
        CMFT \cite{he2023CMFT} & \ding{56}& 94.4 & 90.1 & 94.5&92.2 \\

      \hline
       YOLOv4 \cite{yolov4}&\ding{52}&73.8&79.4&77.9&76.5\\
       YOLOv7 \cite{yolov7}&\ding{52}&81.9&83.8&86.1&82.8\\
       YOLOv3 \cite{redmon2018yolov3}&\ding{52}&89.2&84.5&88.8&86.8\\
       YOLOv5n \cite{yolov5}&\ding{52}&90.0&85.9&89.3&87.9\\
  $\diamond$CenterNet \cite{duan2019centernet} & \ding{52}&93.5 & 90.3 & 92.1 & 91.9 \\ 
$\diamond$FCOS \cite{tian2019fcos}&\ding{52}&94.1&90.4&92.8&92.2\\ 
Cascade RCNN \cite{cai2019cascadercnn}& \ding{52} & 94.3 & 87.6 & 92.7 &90.8\\
        Faster RCNN \cite{ren2016fasterrcnn} & \ding{52}& 95.1 & 82.5 & 93.1&88.4 \\ 
        RetinaNet \cite{lin2017focal} & \ding{52}& 94.9 & 86.5 & 93.4 &90.5\\     
       ConsistencyDet \cite{jiang2024consistencydet}&\ding{52}&93.1&90.3&94.2&91.7\\    
       DiffusionDet \cite{chen2023diffusiondet}&\ding{52}&94.9&92.1&95.0&93.5\\

            DiffDet4SAR \cite{zhou2024diffdet4sar} & \ding{52}& \underline{95.0}& \underline{92.1} & \underline{95.1}& \underline{93.5}\\
        MaDiNet(Ours) & \ding{52}&\textbf{95.6} &\textbf{96.2}& \textbf{97.5}&\textbf{95.9}\\

       \bottomrule
    \end{tabular}
\end{table}

\begin{table*}[!ht]  

\setlength{\tabcolsep}{2.5pt}
    \centering
    \caption{Comparisons of detection performance on SAR-AIRcraft1.0 dataset \cite{zhirui2023sarcraft}. Results of each category are computed by mAP$_{50}$ (\%) and mAP$_{75}$ (\%). The \textbf{best} and \underline{second best} results are shown in \textbf{boldface} and \underline{underline} (confidence=0.5).$
\diamond$ represents anchor-free methods. }\label{table_aircraft_results}
    \scalebox{1}{
    \begin{tabular}{ccccccccccccccccc}
    
\toprule
        \multirow{2}{*}{Methods} & open- & \multirow{2}{*}{mAP$_{50}$} & \multicolumn{2}{c}{A330}  &  \multicolumn{2}{c}{A320/321} & \multicolumn{2}{c}{A220} & \multicolumn{2}{c}{ARJ21}& \multicolumn{2}{c}{Boeing737} & \multicolumn{2}{c}{Boeing787} & \multicolumn{2}{c}{other} \\ 
        \cline{4-17}
        ~ & source & ~ & AP$_{50}$ & AP$_{75}$ & AP$_{50}$ & AP$_{75}$ & AP$_{50}$ & AP$_{75}$ & AP$_{50}$ & AP$_{75}$ & AP$_{50}$ & AP$_{75}$ & AP$_{50}$ & AP$_{75}$ & AP$_{50}$ & AP$_{75}$ \\ \midrule
       $\diamond$ SKG-Net \cite{fu2021SKGNET} & \ding{56} & 70.5 & 79.3 & 66.4 & 78.2 & 49.6 & 66.4 & 29.8 & 65.0 & 37.7 & 65.1 & 48.7 & 69.6 & 51.6 & 71.4 & 41.4 \\ 
        SA-Net \cite{zhirui2023sarcraft} & \ding{56} & 77.7 & 88.6 & 88.6 & 94.3 & 86.6 & 90.3 & 55.0 & 78.6 & 59.7 & 59.7 & 41.8 & 70.8 & 60.4 & 71.3 & 47.7 \\ 
        MLSDNet \cite{chang2023mlsdnet} & \ding{56}&82.7& 91.5 &-& 96.9 &-& 85.1 &-& 83.2 &-& 71.7&- & 72.1 &-& 78.4 &-\\

 \hline

       $\diamond$ FCOS \cite{tian2019fcos} & \ding{52} & 54.8 & 30.5 & 29.4 & 65.1 & 63.7 & 59.7 & 32.5 & 56.6 & 35.1 & 41.4 & 19.2 & 45.8 & 34.1 & 62.3 & 32.5 \\ 
       $\diamond$ CenterNet \cite{duan2019centernet} & \ding{52} & 70.5 & 90.8 & 68.7 & 91.7 & 64.0 & 70.1 & 43.5 & 63.6 & 45.1 & 47.0 & 25.4 & 65.8 & 49.1 & 65.1 & 40.4 \\
        RetinaNet \cite{lin2017focal} & \ding{52} & 72.3 & 92.0 & 70.1 & 92.6 & 58.4 & 73.0 & 41.7 & 63.2 & 47.1 & 47.8 & 25.3 & 65.4 & 50.0 & 67.0 & 42.3 \\ 
       $\diamond$ RepPoints \cite{yang2019reppoints} & \ding{52} & 72.6 & 89.8 & 66.4 & 97.9 & 84.9 & 71.4 & 49.4 & 73.0 & 50.9 & 55.7 & 36.6 & 51.8 & 41.8 & 68.4 & 43.1 \\ 
        Cascade RCNN \cite{cai2019cascadercnn} & \ding{52} & 75.4 & 87.1 & 87.2 & 97.4 & 73.5 & 73.5 & 58.0 & 77.4 & 59.0 & 54.1 & 39.0 & 68.1 & 57.2 & 68.4 & 45.6 \\ 
        Faster RCNN \cite{ren2016fasterrcnn} & \ding{52} & 75.9 & 84.8 & 84.8 & 96.7 & 87.2 & 78.1 & \underline{58.2} & 73.4 & 54.3 & 54.5 & 42.4 & 72.5 & 60.1 & 69.4 & 44.9 \\ 
        YOLOX \cite{ge2021yolox} & \ding{52} & 81.0 & 95.2 & 74.4 & 96.7 & 74.4 & 79.3 & 45.0 & 78.2 & 39.2 & 66.4 & 39.5 & 78.0 & 50.7 & 73.2 & 37.5 \\ 
        YOLOv3 \cite{redmon2018yolov3} & \ding{52} & 83.9 & 91.8 & 90.8 & 96.9 & \textbf{97.0} & \underline{86.5} & \textbf{69.6} & 77.5 & \underline{61.4} & \underline{77.0} & 52.6 & 76.4 & 65.8 & 82.4 & 57.2 \\ 
ConsistencyDet \cite{jiang2024consistencydet}&\ding{52}& 85.1& 93.4&80.2&96.1&81.0&78.8&54.2&81.2&58.6&68.2&58.1&89.4&75.0&83.4&45.7   \\
        DiffusionDet \cite{chen2023diffusiondet}&\ding{52}& 86.6& 95.4&88.3&98.1&83.0&80.8&55.0&84.2&59.7&70.9&60.1&91.4&77.0& \underline{86.4}&48.6   \\
        YOLOv5n \cite{yolov5} & \ding{52} & 88.2 & 88.2 & 83.3 & 98.9 & 68.2 & 84.6 & 52.5 & 86.6 & 56.1 & 75.0 & \underline{69.3} & \textbf{95.2} & 77.6 & 84.7 & 54.4 \\ 
        YOLOv8n \cite{yolov8} & \ding{52} & 88.2 & 93.0 & \underline{92.0} & 97.0 & 72.9 & 85.4 & 56.1 & 86.0 & \textbf{66.1} & 74.5 & \textbf{70.3} & 91.0 & \textbf{82.3} & 83.0 & \textbf{58.0} \\

        DiffDet4SAR \cite{zhou2024diffdet4sar} & \ding{52} & \underline{88.4} & \underline{97.1} & 89.6 & \underline{99.4} & 88.4 & 82.3 & 50.6 & \underline{87.2} & 56.0 & 72.8 & 64.7 & 93.3 & 77.1 & 85.9 & 49.7\\
        MaDiNet(Ours) &   \ding{52} & \textbf{90.5} & \textbf{98.7} & \textbf{96.5} & \textbf{99.6} & \underline{96.5} & \textbf{86.9} & 54.8 & \textbf{89.2} & 56.9 & \textbf{77.1} & 67.6 & \underline{93.9} & \underline{80.2} & \textbf{87.9} & \underline{57.8}\\
 \bottomrule
    \end{tabular}
}
\end{table*}

\textbf{2) SAR-ShipDatatset.} The SAR-ShipDataset\footnote{download link: \url{https://github.com/CAESAR-Radi/SAR-Ship-Datasetcomprises}} is the largest publicly available SAR ship detection dataset, which is derived from 102 images captured by the Chinese Gaofen-3 satellite and 108 images from Sentinel-1 \cite{wang2019sarshipdataset}. The images are acquired in various imaging modes and resolutions and include scenes such as harbors, reefs, coastal waters, and sea surfaces with varying sea clutter intensities. The image patches are of size 256×256 pixels. The revised dataset comprises a total of 39,729 images and 47,416 ship instances.

\textbf{3) SAR-AIRcraft1.0.} The SAR-AIRcraft1.0\footnote{download link: \url{https://radars.ac.cn/web/data/getData?newsColumnId=f896637b-af23-4209-8bcc-9320fceaba19}} is the first aircraft detection dataset containing fine-grained recognition labels, which is collected from the Gaofen-3 satellite, with a resolution of 1 meter and a spotlight imaging mode, using single-polarization \cite{zhirui2023sarcraft}. It encompasses seven categories, including A220, A320/321, A330, ARJ21, Boeing 737, Boeing 787, and \textit{other}, where \textit{other} denotes aircraft instances that do not belong to the other six categories. The dataset contains multi-temporal images from different airports, covering a large area with complex backgrounds. The dataset includes images of four different sizes: 800×800, 1000×1000, 1200×1200, and 1500×1500, totaling 4,368 images with 16,463 aircraft instances.

\subsection{Implementation details}
 We compare two dozen advanced detection models on three datasets to verify the effectiveness of the proposed method, including traditional, anchor-based, and anchor-free methods. To increase the reproducibility of the article, the open-source and non-open-source methods are also marked separately. All hyperparameters of open-source compared methods follow the settings mentioned in their papers. Our network was optimized by the AdamW method with an initial learning rate of $ 1.5 \times 10^{-5}$, which was decayed by $1 \times 10^{-4}$. All the experiments were implemented in PyTorch on a computer with an NVIDIA RTX 4090 and 24GB memory. More details of the experimental setup can be found on our GitHub\footnote{ \url{https://github.com/JoyeZLearning/MaDiNet}}.

\begin{figure}
	\centering
	\includegraphics[width=0.9\columnwidth]
{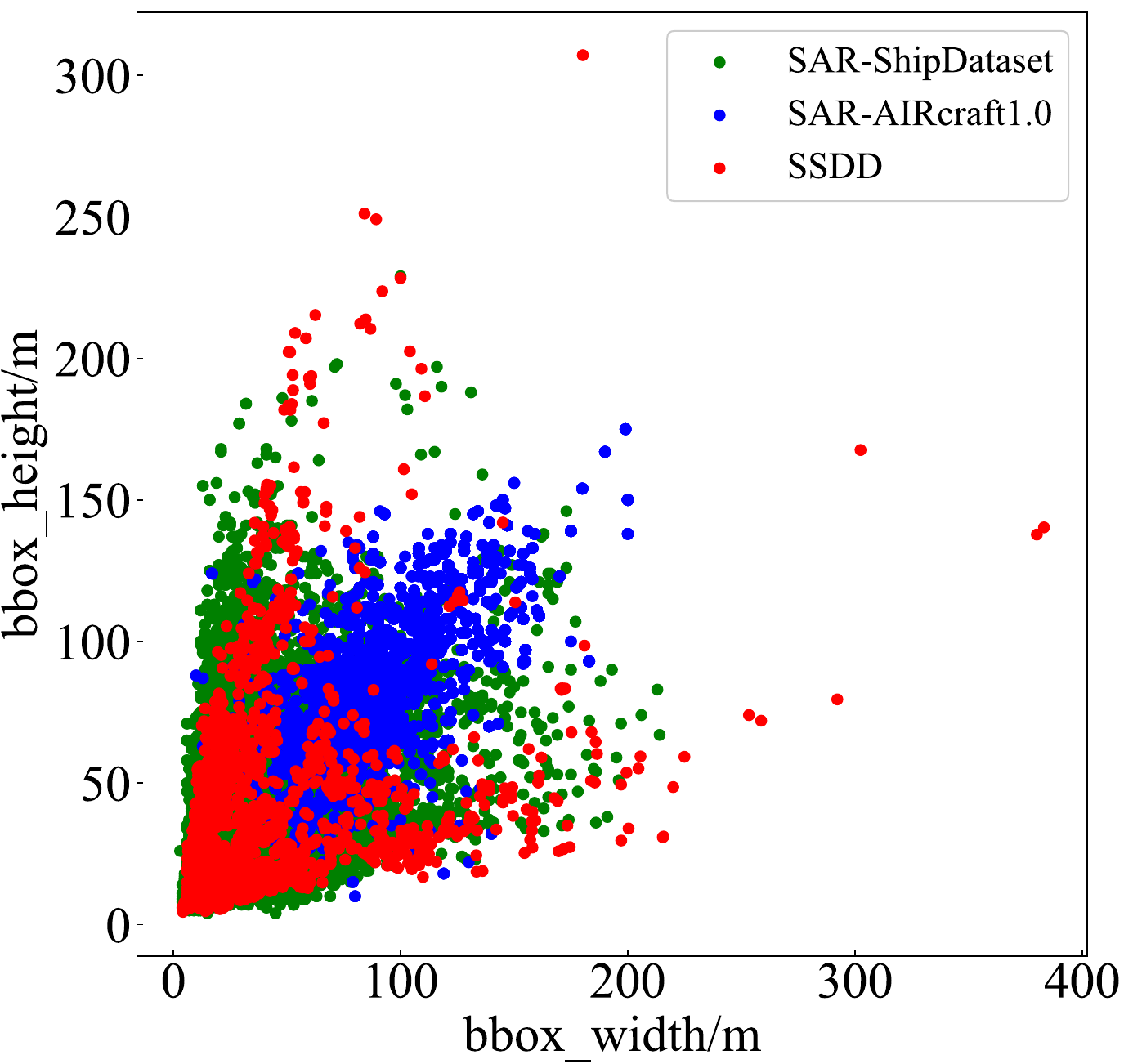}
	\caption{Bounding box sizes of targets in three datasets employed in our experiments, and bbox is an abbreviation for bounding boxes (SAR-ShipDataset \cite{wang2019sarshipdataset}, SAR-AIRcraft1.0 \cite{zhirui2023sarcraft}), SAR-AIRcraft1.0 \cite{zhirui2023sarcraft}). }
	\label{bboxSize}
\end{figure}

\subsection{Evaluation Metrics}

To evaluate the performance of different methods, we employ four key evaluation metrics:  precision (P), recall (R), F1-score, and mean average precision (mAP). P represents the actual number of positive samples in the predicted positive sample. R represents the proportion of positive samples that are judged to be positive. And mAP represents the average precision values for multiple classes.

\begin{figure*}[t]
	\centering
	\includegraphics[width=2\columnwidth]
{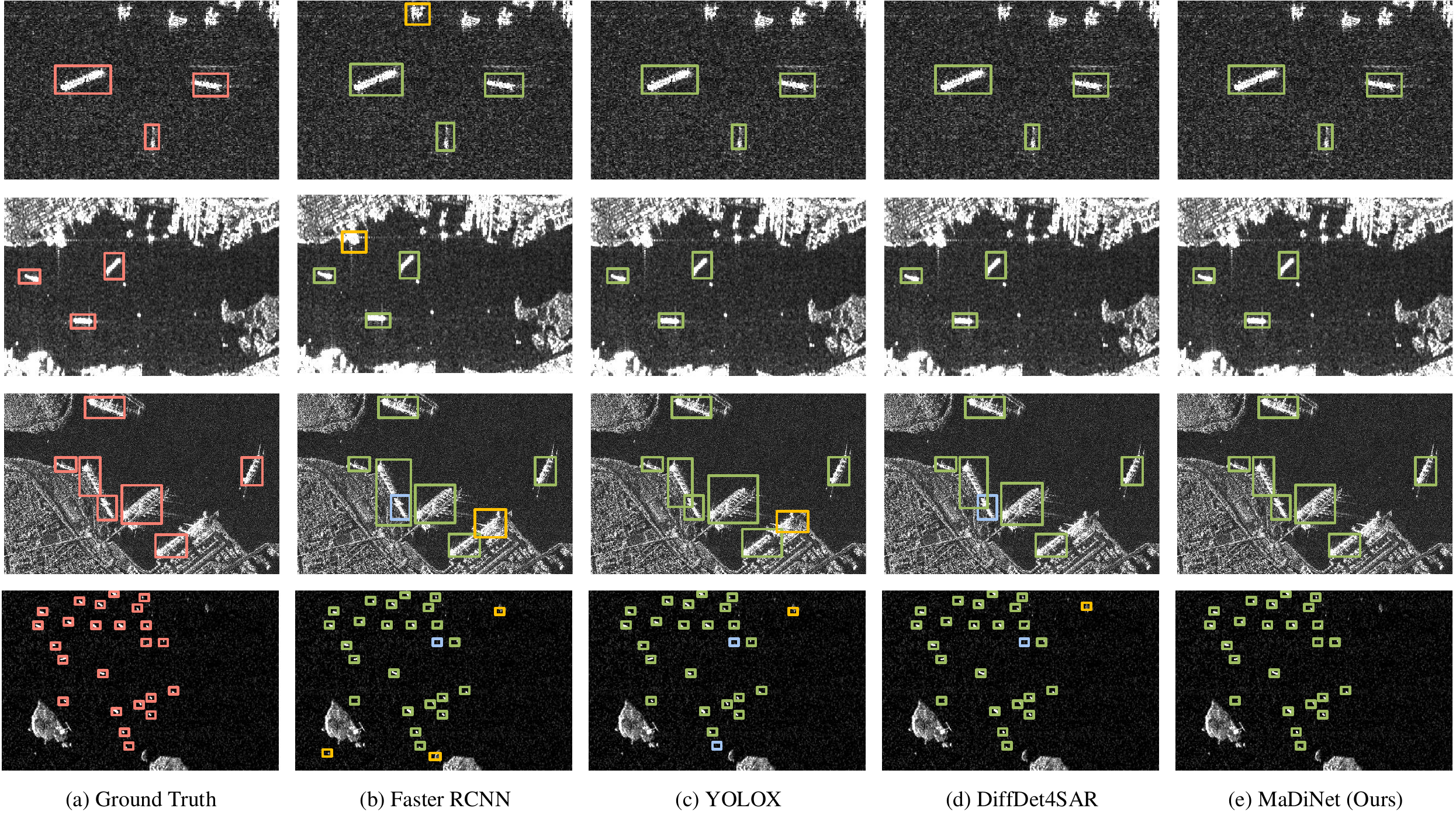}
	\caption{Visualization of detection results under different methods on  SSDD \cite{zhang2021sarssdd}. Red boxes represent ground truth, and green boxes indicate detection results. Orange boxes indicate false positives, and the blue boxes represent false negatives.}
	\label{visualization_SSDDResults}
\end{figure*}

\begin{figure*}
	\centering
	\includegraphics[width=2\columnwidth]
{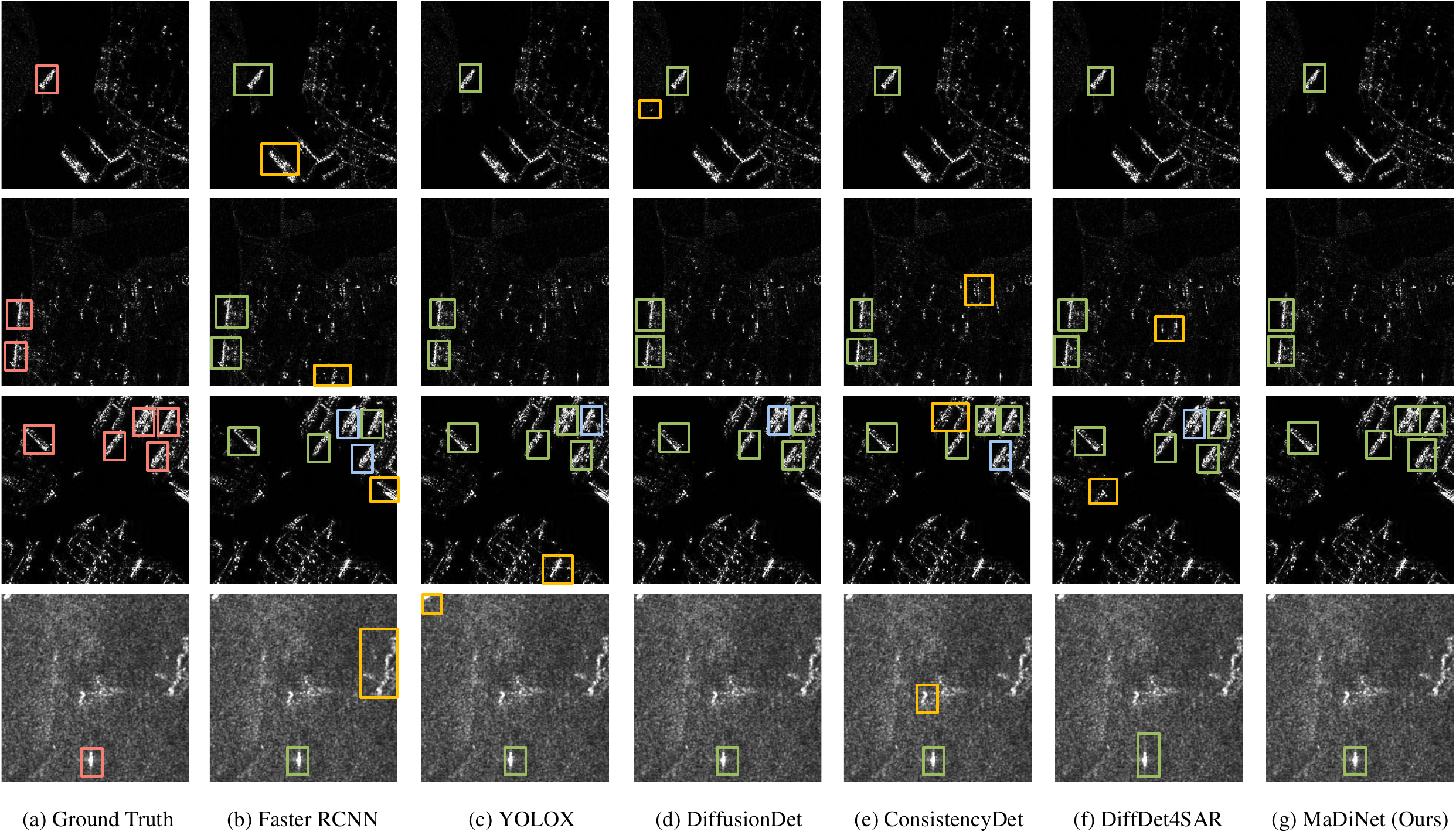}
	\caption{Visualization of detection results under different methods on SAR-ShipDataset \cite{wang2019sarshipdataset}. Red boxes represent ground truth, and green boxes indicate detection results. Orange boxes indicate false positives, and the blue boxes represent false negatives.}
	\label{visualization_sarshipdatasetResults}
\end{figure*}

\begin{figure*}
	\centering
	\includegraphics[width=2\columnwidth]
{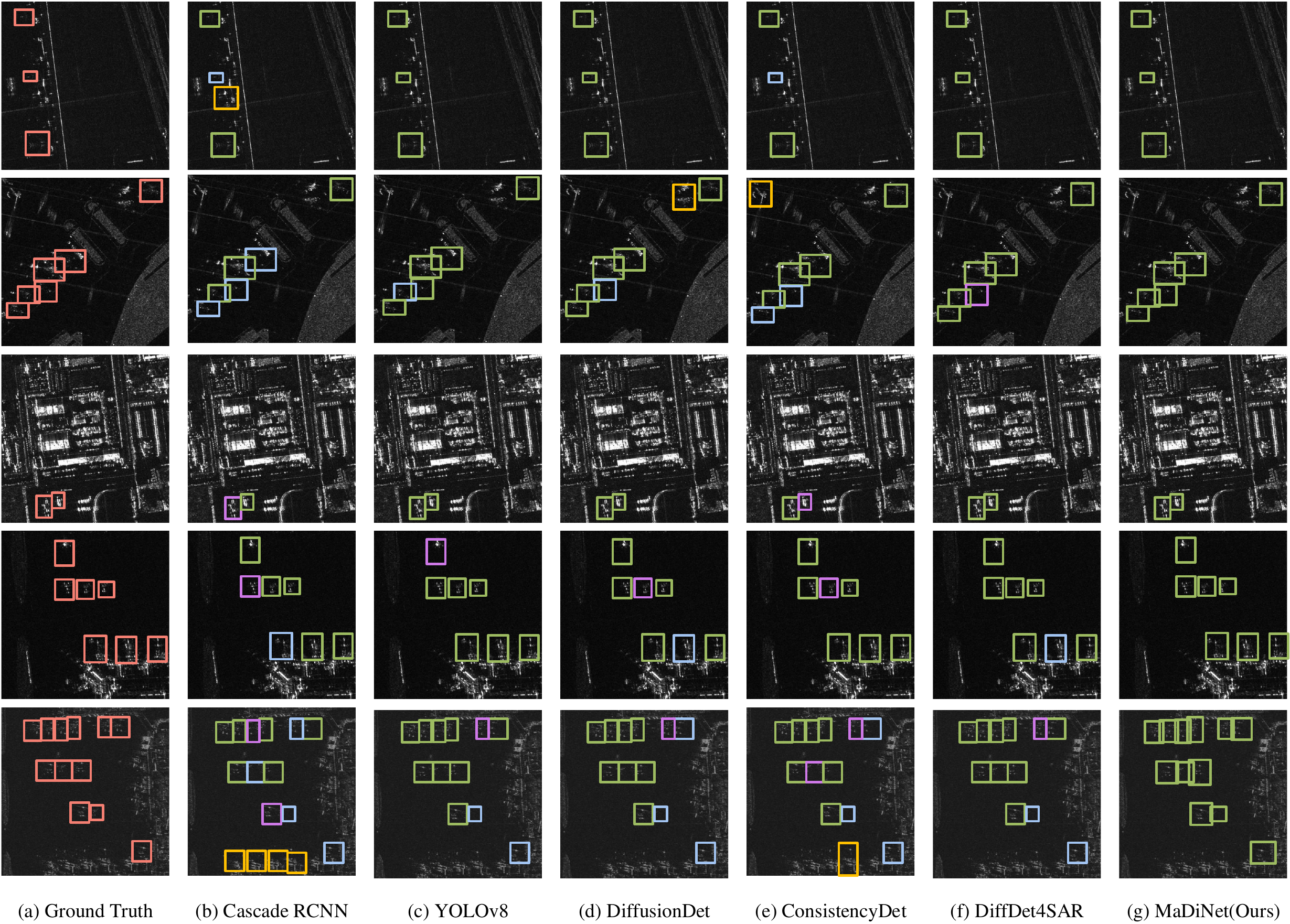}
	\caption{Visualization of detection results under different methods on SAR-AIRcraft1.0 dataset \cite{zhirui2023sarcraft}. Red boxes represent ground truth, and green boxes indicate detection results. Orange, blue, and purple boxes indicate false positives, false negatives, and misidentified aircraft types, respectively.}
	\label{visualization_aircraftResults}
\end{figure*}

\begin{figure*}
\centering
	\begin{minipage}{0.22\linewidth}
 \centering
		\includegraphics[width=\textwidth]{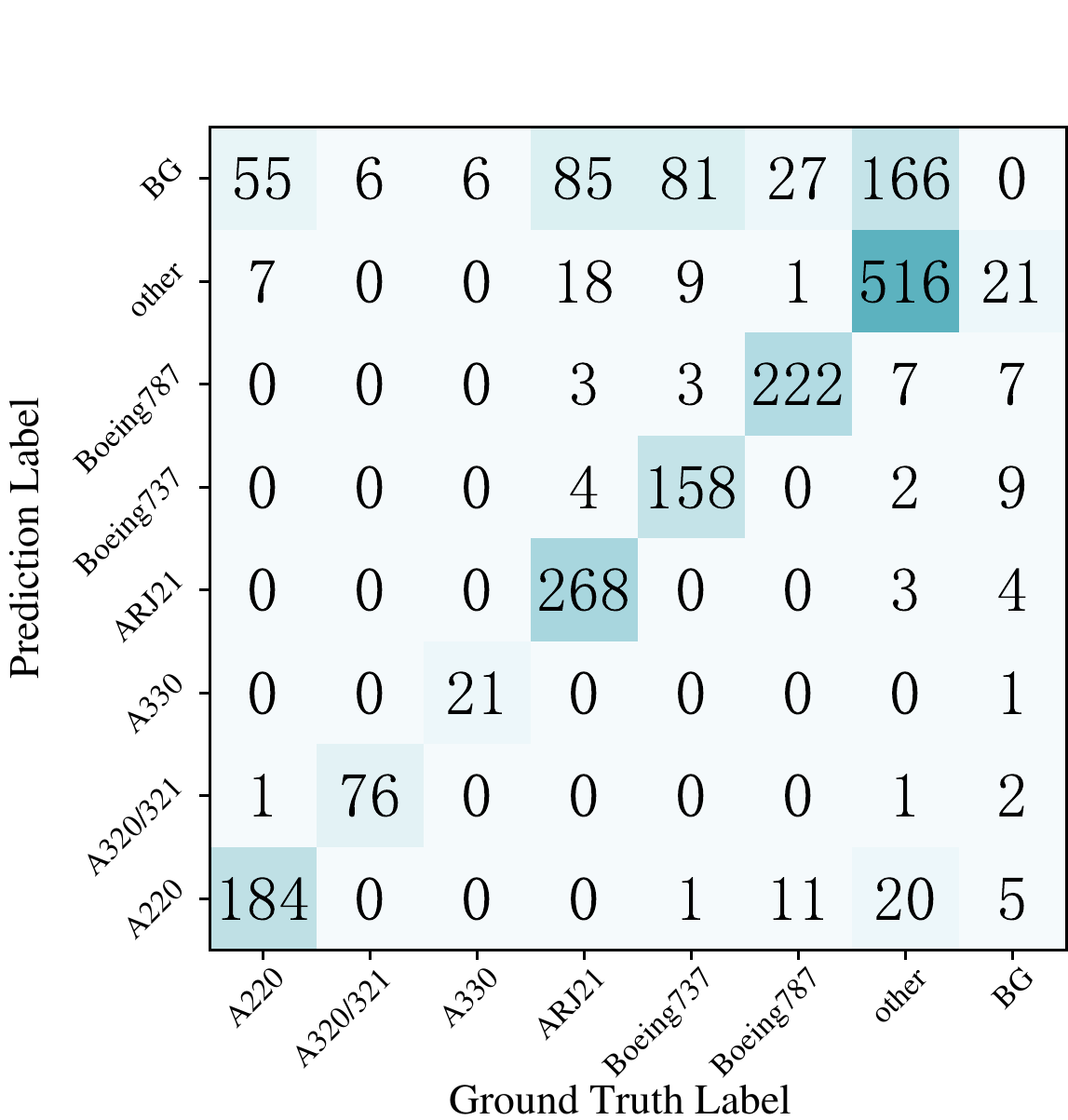}
		\centerline{(a) ConsistencyDet}
	\end{minipage}
	\begin{minipage}{0.22\linewidth}
  \centering
		\centerline{\includegraphics[width=\textwidth]{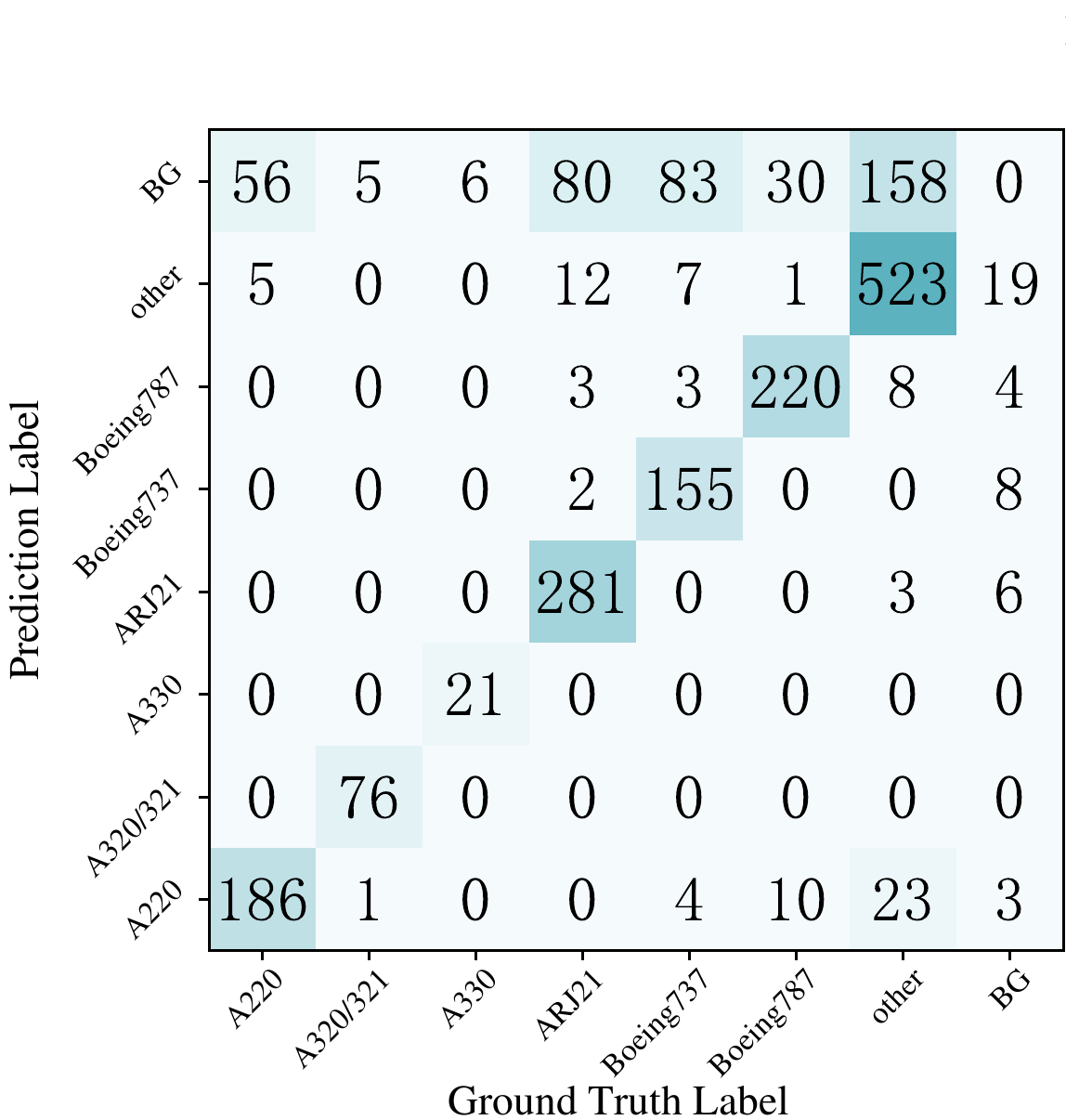}}
		\centerline{(b) DiffusionDet}
	\end{minipage}
	\begin{minipage}{0.22\linewidth}	
  \centering
		\centerline{\includegraphics[width=\textwidth]{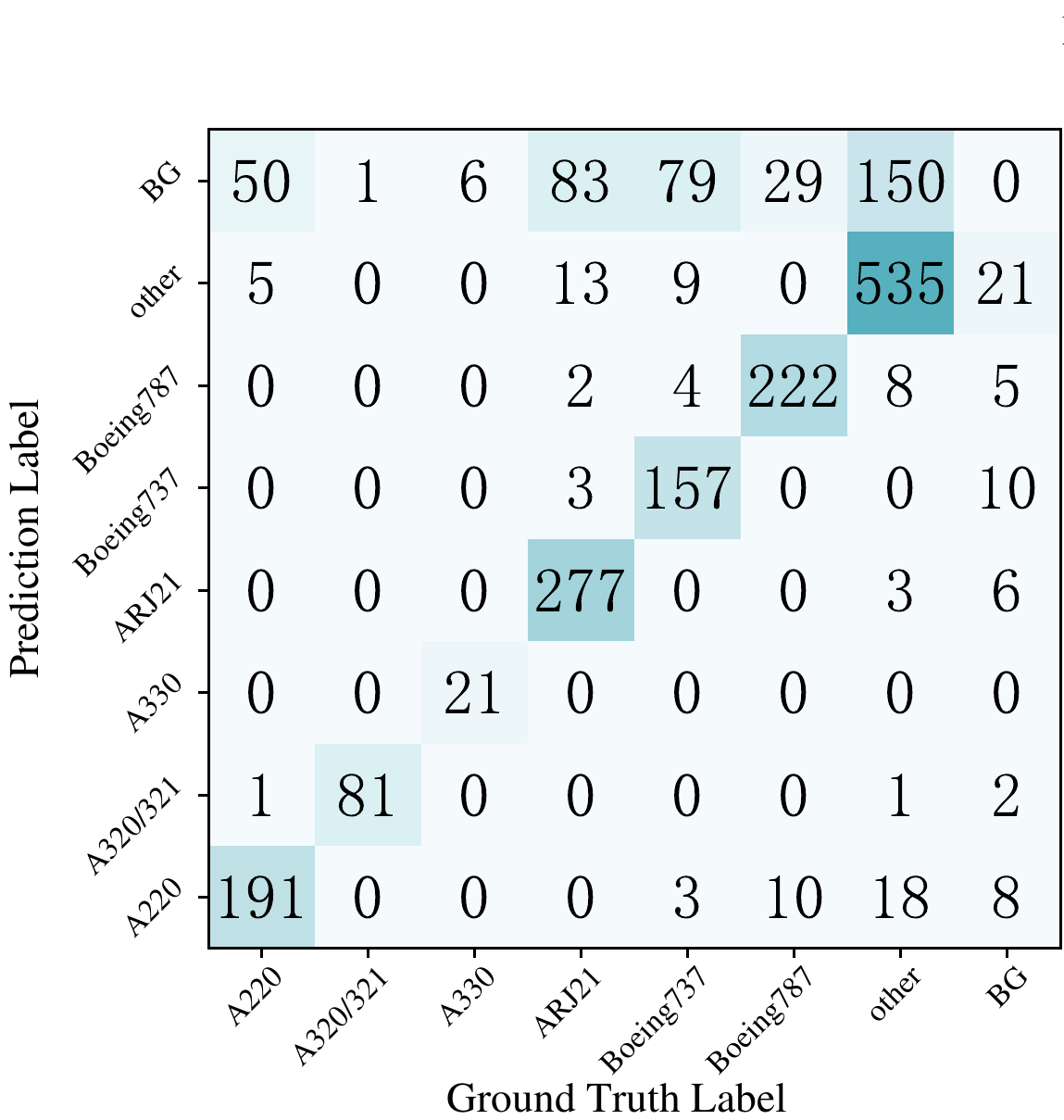}}	 
		\centerline{(c) DiffDet4SAR}
	\end{minipage}
 	\begin{minipage}{0.2578\linewidth}
   \centering
		\centerline{\includegraphics[width=\textwidth]{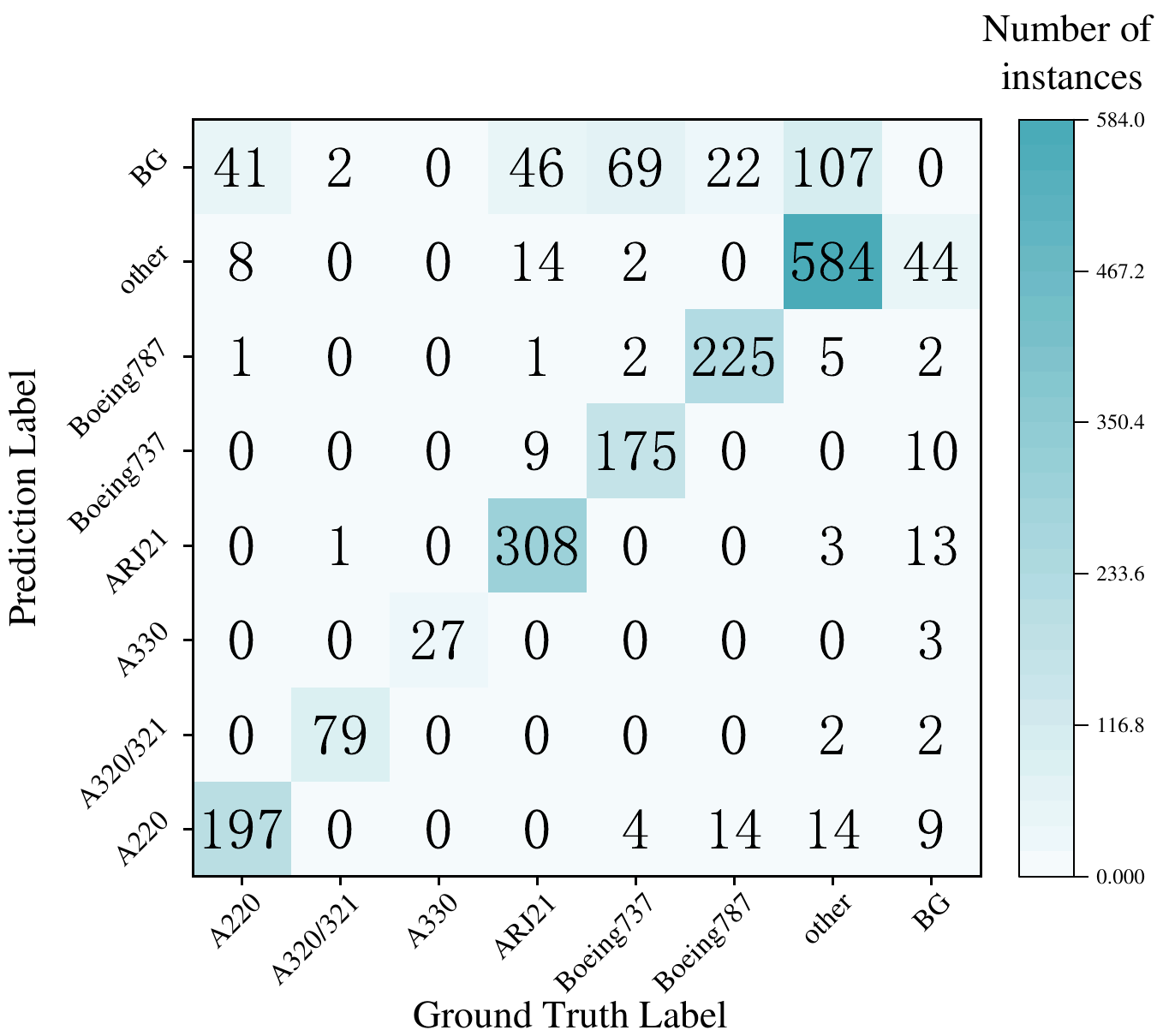}}	 
		\centerline{(d) MaDiNet (Ours)}
	\end{minipage}
 
 \caption{Confusion matrices of different methods using SAR-AIRcraft1.0 dataset \cite{zhirui2023sarcraft}. BG represents background.}
	 \label{fig:confusionMatrix} 
\end{figure*}

\section{Experimental results and discussion}\label{experimentresults}

\subsection{Quantitative Results}
Fig. \ref{bboxSize} illustrates the size of targets in various datasets.  It is evident that the aspect ratio of ship targets significantly exceeds that of aircraft targets, and there exists a wide scale range for similar targets across each dataset. Table \ref{tabel_ssdd_results}, Table \ref{table_shipdata_results}, and Table \ref{table_aircraft_results} present a quantitative comparison of the proposed method against two dozen state-of-the-art target detection models on the SSDD \cite{zhang2021sarssdd}, SAR-ShipDataset \cite{wang2019sarshipdataset} and SAR-AIRcraft1.0 dataset \cite{zhirui2023sarcraft}, respectively. The results indicate that our method achieves superior detection performance compared to all other methods when confronted with a diverse target scale and a wide range of target scales. On SSDD, the proposed MaDiNet achieves a mAP$_{50}$ of 99.0\%. For the SAR-ShipDataset, the proposed method demonstrates the highest mAP$_{50}$ value of 97.5\%, surpassing the second-method method DiffDet4SAR by 2.4\%. On the SAR-Aircraft-1.0 dataset, the proposed MaDiNet sets a new state-of-the-art with a mAP$_{50}$ of 90.5\%, outperforming DiffDet4SAR by 2.1\% . In additional, MaDiNet showcases superior fine-grained recognition performance, achieving top 
detection accuracy for four target categories. Among them, mAP$_{75}$ on both A330 and A320/321 is as high as 96.5\%, validating the localization and recognition capabilities of the proposed method. We attribute this to two reasons. On the one hand, we use the diffusion process for random sampling and denoising of the anchor boxes to achieve flexible positioning and improve the recall rate, which can better adapt to the multi-scale characteristics of the target in SAR images. On the other hand, due to the global context information modeling of designed MambaSAR, background interference is suppressed and target feature representation is enhanced, which is beneficial for SAR image target detection under complex background.

\subsection{Qualitative Results}
We also present the visualization results of different methods on three datasets.
 
\textbf{SSDD.} To visualize and compare detection results, we take four classical complex images from various scenes, as Fig. \ref{visualization_SSDDResults} illustrates. For near-shore dense ship scenarios, Faster R-CNN frequently misidentifies metal buildings that are not man-made as targets, which can lead to false alarms. Alternatively, MaDiNet can detect ships that are irregularly distributed at different scales. In low-resolution large-scale contexts, MaDiNet not only accurately detects near-shore ships but also effectively recognizes densely packed small-scale vessels in coastal ports. The improved performance can be ascribed to the design of the MambaSAR module, which enhances target saliency by capturing richer features.

\textbf{SAR-ShipDataset.} As Fig. \ref{visualization_sarshipdatasetResults} illustrates, Faster R-CNN and ConsistencyDet show a large number of false alarms in both near-shore and offshore scenarios. This problem results from azimuthal blurring and pseudo-signals produced by metal buildings in the near-shore environment. These buildings and ships have similar scattering characteristics, which confuses the model. MaDiNet, in comparison, achieves better detection results with fewer false alarms and missed detections. We attribute this to the incorporation of an attention-based mechanism within the MambaSAR backend. This enhancement enables the detector to more effectively filter important information based on input information, thereby enhancing its ability to capture global semantics. Moreover, our method is more robust to dynamic clutters in the scene, which also benefited from the MambaSAR module. 

\textbf{SAR-AIRcraft1.0.} As seen in Fig. \ref{visualization_aircraftResults}, we chose five representative images of intricate scenes to compare visually. It can be observed that Cascade R-CNN and ConsistencyDet generate numerous false negatives and misjudgments. This is due to the strong scattering points of surrounding background features, like towers and terminal buildings, which can weaken target scattering and confuse parts of the aircraft with these background features, making it challenging to distinguish and locate the aircraft correctly. In contrast, in environments with strong background interference, our suggested MaDiNet performs noticeably better than other methods. This advantage stems from our implementation of random noise bounding boxes. In contrast to conventional fixed bounding boxes, the diffusion-based detector can more effectively adjust to multi-scale variations of targets because it is not limited by the size or aspect ratio of anchors. More importantly, this performance enhancement is attributed to our design of four scanning mechanisms with symmetric branches without SAR-SSM within the MambaSAR module. With systematic orientation transformations, this layout improves both the efficiency and the comprehensiveness of capturing image features by covering every region of the input image and offering a rich multidimensional information base for subsequent feature extraction.

To further quantitatively assess model performance and provide more insights into the detection results, confusion matrices for fine-grained recognition across three distinct diffusion-based detectors and our proposed MaDiNet are shown in Fig. \ref{fig:confusionMatrix}. Notably, values are primarily concentrated along the diagonal line in the proposed MaDiNet confusion matrix, indicating a low misclassification rate and demonstrating the effectiveness of our method in correctly identifying targets. This is attributed to the designed mambaSAR module. The MambaSAR module models the interaction between the target and its environment interactively, greatly enhancing the feature representation of targets. When compared to other categories, the A330 and A320/321 categories consistently show superior recognition results. This is mainly because the A320/321 aircraft have unique sizes that make them easier to recognize. Their fuselage length is more than 40 meters. In contrast, due to their smaller size, which makes it more difficult to capture detailed features, some target types, such as A220 and ARJ21 aircraft, show lower detection accuracy.

\begin{table}[!ht]
     \centering
     \caption{Ablation study of different backbone in MaDiNet on three different datasets. $\star$ denotes MaDiNet without the MambaSAR module. The best results are shown in \textbf{bold} (mAP$_{50}$/\%, confidence=0.5).}\label{number_backbone_results}  
 \tabcolsep=5pt

    \begin{tabular}{cccc}
    \toprule
       \multirow{2}{*}{Backbone} & \multirow{2}{*}{SSDD} & SAR- &SAR-\\
       & &  AIRcraft1.0 &ShipDataset \\ \midrule 
        Res50 & 95.2 & 85.0 & 93.5 \\ 
        Res50-FPN & 96.9 & 88.4 & 95.1 \\ 
        Res50-BiFPN & 85.7 & 73.0 & 85.7 \\ 
        Res101 & 92.3 & 81.0 & 90.3 \\ 
        ConvNeXt & 95.6 & 85.7 & 94.1 \\ 
        Swin-Transformer & 78.1 & 61.5 & 79.2 \\ 
        MambaVision & 95.3 &86.8 &96.1 \\
        Ours$\star$ & 97.8 & 88.9 & 96.2 \\ 
        \rowcolor{pink} Ours & \textbf{99.0} & \textbf{90.5} & \textbf{97.5} \\ \bottomrule
    \end{tabular}
\end{table} 

\subsection{Ablation Study}
In this subsection, we conduct different ablation studies to investigate the design of MaDiNet.
\subsubsection{The effectiveness of the backbone design}
We present the experimental results in Table \ref{number_backbone_results}, maintaining the overall architecture of the proposed MaDiNet while comparing various backbones. The traditional Resnet50 serves as our baseline. It can be observed that our proposed backbone outperforms Resnet50 by 3.8\%, 5.5\%, and 4\% in mAP$_{50}$ on three datasets, respectively. This is attributed to the hybrid architecture design combining local modeling from CNN and global modeling from Mamba, effectively enhancing the feature extraction capability of the model. Specifically, we also assess MambaVision, a pure Mamba structure, as a baseline. Nevertheless, rather than improving performance, this structure causes it to decline. The decline in SAR images can be ascribed to the presence of complex backgrounds and significant scattering noise. As a state linear space model, Mamba prioritizes the correlation of global information, but during the extraction process, it unintentionally incorporates these intricate background features, causing issues with detection precision. In response, we incorporate convolution operations into a state space model with dynamic weighting to design the MambaRes50-FPN backbone. This integrated backbone effectively enhances detection performance across various target types by leveraging the strengths of global feature modeling inherent in Mamba.

\begin{table}[!ht]
    \centering 
 \tabcolsep=0.5pt
    \caption{Influence of the location of MambaSAR module on MaDiNet using SAR-AIRcraft1.0 dataset \cite{zhirui2023sarcraft}. MambaSAR$^{\ddagger}$ denotes the MambaSAR module without downsampling. The best results are shown in \textbf{bold} (mAP$_{50}$/\%, confidence=0.5).}
    \label{position_mamba_backbone_results}
    \begin{tabular}{cc}  
    \hline
        Backbone & mAP$_{50}$ \\ \hline
        Res50\_FPN & 88.4  \\ 
 Res3=MambaSAR(Res2) & 88.5 \\ 
          Res4=MambaSAR(Res3) & \underline{89.8}  \\
            Res5=MambaSAR(Res4) & 87.5  \\       
        Res4=MambaSAR$^{\ddagger}$(Res4) & 88.0  \\ 
         Res5=MambaSAR$^{\ddagger}$(Res5) & 86.8  \\ 
        Res5=MambaSAR$^{\ddagger}$(Res5), Res4=MambaSAR$^{\ddagger}$(Res4) & 88.2  \\ 
        Res5=ori\_res5+MambaSAR(Res4) & 87.8\\ 

         \rowcolor{pink} Res4=ori\_Res4+MambaSAR(Res3) & \textbf{90.5}  \\ \hline
    \end{tabular}
\end{table}

\begin{figure}
	\centering
	\includegraphics[width=1\columnwidth]
{position_mambasar.pdf}
	\caption{Schematic diagrams of the position of the MambaSAR module. MambaSAR$^{\ddagger}$ denotes the MambaSAR module without downsampling. \newline
 (a) Res3=MambaSAR(Res2). (b) Res4=MambaSAR(Res3).
 (c) Res5=MambaSAR(Res4). (d) Res5=MambaSAR$^{\ddagger}$(Res5). 
 (e) Res4=ori\_Res4+MambaSAR(Res3).\newline
 (f) Res5=ori\_Res5+MambaSAR(Res4). \newline
 (g) Res5=MambaSAR$^{\ddagger}$(Res5), Res4=MambaSAR$^{\ddagger}$(Res4). \newline
 (h) Res4=MambaSAR$^{\ddagger}$(Res4).}
	\label{position_mambasar}
\end{figure}

\begin{table*}[!ht]
    \centering
    \caption{Ablation study on the effectiveness of different hybrid  patterns within MambaSAR module using SAR-AIRcraft1.0 dataset \cite{zhirui2023sarcraft}, including the integration sequence of mambablock and attention block and the corresponding number
of layers. $*$ denotes that mambablock is used after the attention layer (AP/\%, confidence=0.5).}
    \label{Influence under different numbers of Mambablock and attention}
    \begin{tabular}{ccccccccc}
    \toprule
        MambaBlock & Attention & Agent-Attention & AP & AP$_{50}$ & AP$_{75}$ & AP$_{s}$ & AP$_{m}$ & AP$_{l}$ \\ \midrule
        - & - & - & 62.3 & 88.9 & 69.0 & 50.0& 60.8 & 55.5 \\ \hline
        1 & - & - & 62.3 & 88.9 & 69.8 & 50.0 & 60.3 & 57.1 \\ 
        2 & - & - & 62.4 & 89.3 & 70.2 & 50.0 & 60.5 & 57.9 \\ 
        3 & - & - & 62.6 & 89.6 & 71.0 & 40.0 & 60.6 & 57.4 \\ \hline
        3 & 1$*$ & - & 62.6 & 89.5 & 70.4 & 40.0 & 61.3 & 57.2 \\ 
        3 & 2$*$ & - & 62.0 & 89.4 & 70.2 &  40.0 &  64.3 & 57.0 \\ 
        3 & 3$*$ & - & 61.9 & 89.1 & 70.0 &  50.0 & 67.7 & 56.9 \\ \hline
        3 & 1 & - & 62.6 & 89.7 & 71.0 & 40.0 & 61.4& 57.4 \\ 
        3 & 2 & - & 62.6 & 89.8 & 70.9 & 40.0 & 61.5 & 57.4 \\ 
        3 & 3 & - & 62.6 & 90.0 & 70.5 & 40.0 & 61.8 & 57.5 \\ \hline
        3 & - & 1 & 63.0 & 89.7 & 70.9 & 40.0 & 61.5 & 57.0 \\ 
        3 & - & 2 & 63.0 & 90.2 & 70.9 & 40.0 & 61.7 & 57.5 \\ 
        \rowcolor{pink} 3 & - & 3 & 63.9 & 90.5 & 70.2 & 40.0 & 62.1 & 57.7 \\ \bottomrule
    \end{tabular}
\end{table*}

\begin{figure*}
	\centering
	\includegraphics[width=2\columnwidth]
{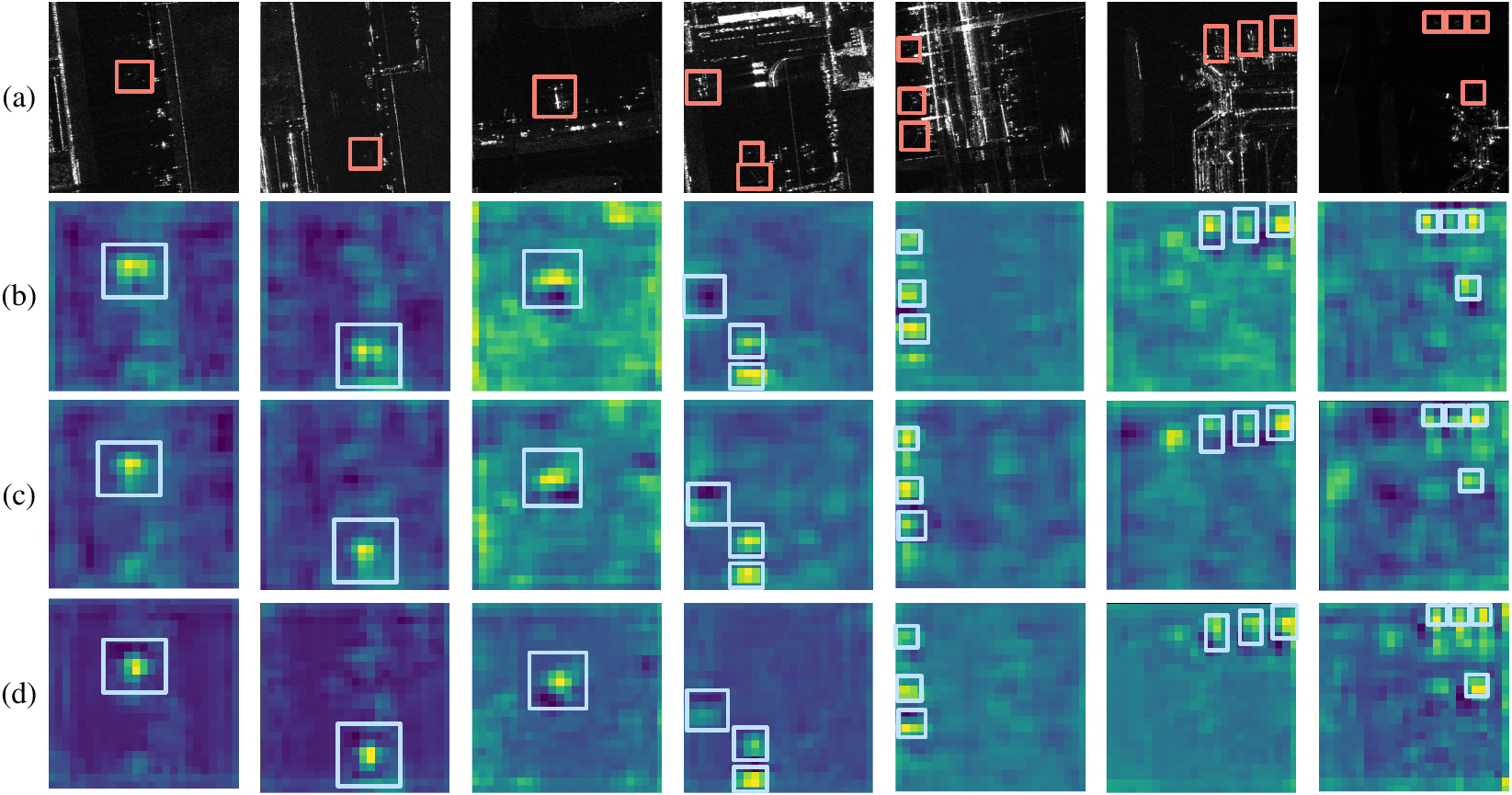}
	\caption{Ablation study on the effectiveness of different hybrid patterns within MambaSAR module using SAR-AIRcraft1.0 dataset \cite{zhirui2023sarcraft}. (a) Ground truth. (b) without MambaSAR module. (c) MambaSAR module without agent attention block. (d) MambaSAR module with agent attention block. After using mambablock, the intensity of the yellow part where the target is located in the feature map is stronger. Boundaries between dense targets are also more obvious. By further adding the agent attention layer, the intensity of the complex background is significantly suppressed and target features are enhanced.}
	\label{mamba_featuremap}
\end{figure*}

\begin{figure}
	\centering
	\includegraphics[width=1\columnwidth]
{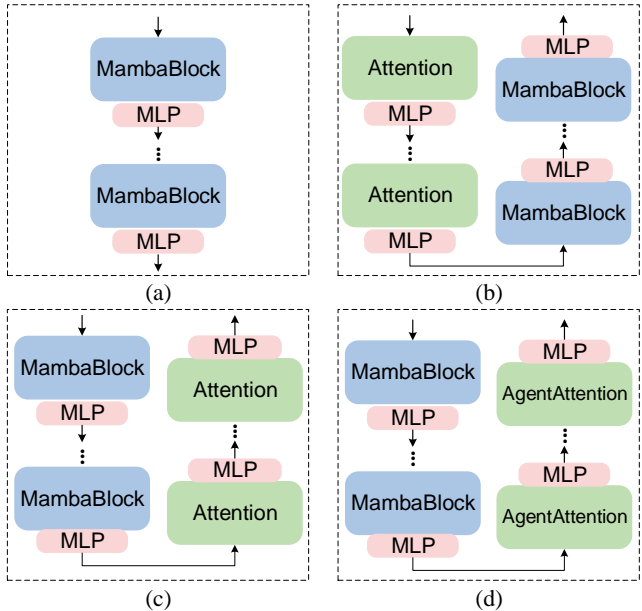}
	\caption{Schematic diagrams of different hybrid patterns. (a) pure Mambablock without an attention layer. (b) Mambablocks are used after attention layers. (c) Mambablocks are used before attention layers. (d) Mambablocks are used before the agent attention layers.}  
	\label{Hybrid_pattern}
\end{figure}

\subsubsection{MambaSAR Module}
In this section, we investigate the effectiveness of the MambaSAR Module, including the position, and the hybrid pattern.

\textbf{Position of MambaSAR module. }We conducted experimental validation of the MambaSAR module position and feature fusion approach on the aircraft dataset (Fig. \ref{position_mambasar}), and the experimental results are shown in Table \ref{position_mamba_backbone_results}. It can be observed that our designed architecture obtains the highest mAP$_{50}$ with 90.5\% on the SAR-AIRcraft1.0 dataset. We then analyze the reasons for our design in detail.

CNNs focus on local features through convolutional kernels, while Mamba filters out irrelevant information using a selective scanning algorithm. Alternatively, it processes rich global information efficiently by storing intermediate results and using kernel fusion, recomputation, and parallel scanning. Feeding Mamba (as shown in Fig. \ref{position_mambasar} (c)(d)(f)(g)(h) the highest layer features extracted by a pure convolutional network directly results in a duplicate set of final features generated by the CNN backbone. On the other hand, since CNNs and Mamba extract features in fundamentally different ways, the Mamba structure struggles to adapt to the abstract high-dimensional features generated by pure CNNs. This mismatch can disrupt the connections between original features. Crucial information is lost before it reaches the detection head because these already compromised high-dimensional semantic features are mapped onto feature maps at all scales via top-down structures like FPN. Consequently, this degradation leads to reduced detection performance. This is also illustrated by the experimental results in  Table \ref{position_mamba_backbone_results}.

Due to the complex scenes of SAR images, using the Mamba structure on shallow features (e.g., Fig. \ref{position_mambasar} (a)) can introduce vast background interference. To address this, we utilize ordinary convolution for feature extraction in the first two layers. This helps filter out unimportant, sparse background information while compacting targets with discrete features. Subsequently, the MambaSAR module is employed to model and dynamically weigh the global sensory fields of low- and medium-dimensional features. As seen in Fig. \ref{position_mambasar} (e), these are then fused with the convolutionally extracted features from the fourth layer to produce new fourth-layer features. It can be observed from Table \ref{position_mamba_backbone_results} that this mode helps the model obtain the better mAP50 of 90.5\%. We attribute this to two reasons. On the one hand, the selective scanning mechanism within the MambaSAR module mitigates local modeling constraints inherent in conventional CNNs, suppresses interference from complex backgrounds, and emphasizes key target features. After passing through a multi-scale fusion mechanism like FPN, critical target information from the fourth layer is mapped across all feature maps at low and medium dimensions. On the other hand, high-dimensional features remain unaffected by Mamba. This architecture ensures effective extraction and fusion of both high-level semantic information and underlying structural details.

\textbf{Hybrid Pattern.} We thoroughly examine in this section how the number of layers and various hybrid integration modes of the mambablock and attention block impact model performance (Fig. \ref{Hybrid_pattern}). We compare the model detection performance when Mambalblocks of different layers are equipped with attention blocks of different types and layers, as indicated in Table \ref{Influence under different numbers of Mambablock and attention}. The model detection performance is shown to improve with an increase in the number of mambablock layers, and the mAP50 of the model with a three-layer mambablock is 0.7\% higher than the model without mamba. The self-attention layer is then used in front of the mambablock. According to experimental findings, utilizing attention before global modeling reduced overall detection accuracy because it was unable to adequately capture the salient characteristics of the target. In contrast, employing the structural sequence of mambablock alongside attention resulted in improved performance, achieving a mAP$_{50}$ of 90.0\%. We contend that target detection accuracy is strongly impacted by the model's comprehension of the scene. Pure Mamba uses linear complexity by compressing filtered global information into hidden states. However, this compression inevitably leads to a loss of fine-grained local dependency information between tokens. As a result, it becomes challenging to perform complete in-context learning and context suppression. In contrast, integrating attention blocks in the later layers of the model helps recover lost global context and captures remote spatial dependencies more effectively. This improvement makes it possible for the model to concentrate on important areas within images and more effectively reduce interference from complicated backgrounds, allowing for finer representations and an improved ability to discern between targets and backgrounds.

Moreover, an additional enhancement in representational capability is obtained by applying an agent-attention layer at the end of each stage, which leads to an instantaneous 0.5\% increase in mAP$_{50}$. This supports our hypothesis that placing attention layers at the end of each stage is beneficial design-wise. Besides, optimal performance was achieved with equal numbers of layers for both attention blocks and mambablocks, resulting in a peak mAP$_{50}$ score of 90.5\%. 

As shown in Fig. \ref{mamba_featuremap}, we visualize and analyze the feature maps before and after integrating the MambaSAR module. After implementing mambablock, we observe a stronger intensity of yellow where the target is located within the feature map. This suggests a stronger emphasis on target characteristics, with more precise localization resulting from distinct boundaries between dense targets. In addition, we observe a significant suppression of complex background intensities when an attention layer is superimposed over this configuration. The model narrows its focus more and more to the ground truth. This observation confirms that our proposed method effectively constrains the model's attention towards genuine SAR target features, thereby improving detection accuracy overall.

\begin{figure*}
	\centering
	\includegraphics[width=2\columnwidth]
{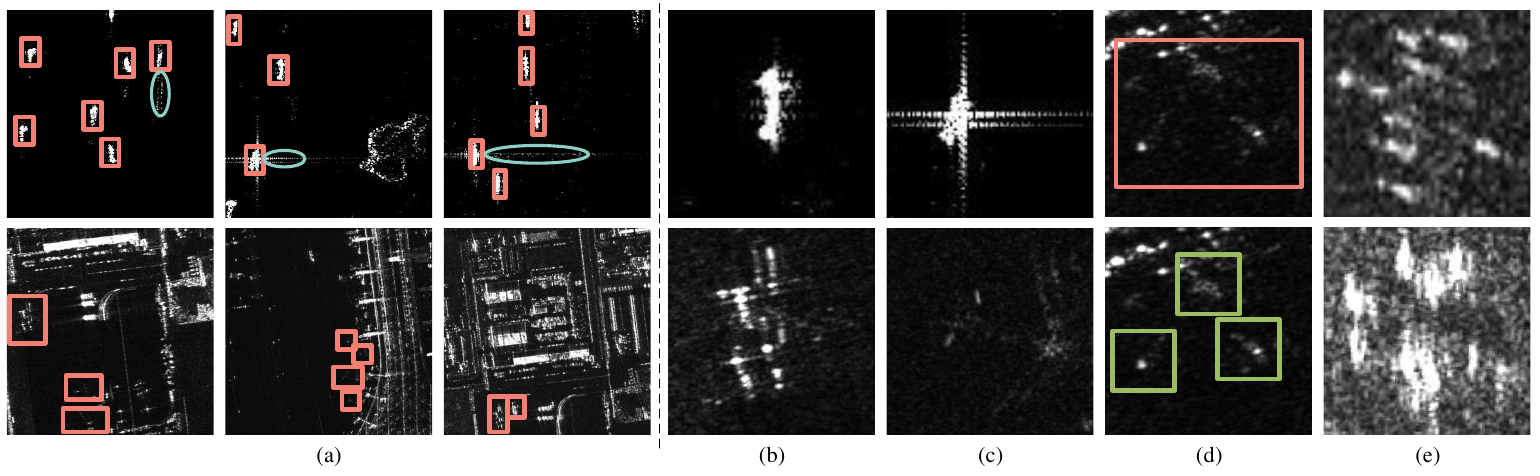}
		\caption{(a) Comparison between the ship dataset and the aircraft dataset for SAR images. The upper row is ships and the lower row is aircraft. Red boxes are the ground truth, and blue ellipse is the wake of ships. (b) (c) Comparison between individual ships and aircraft. The upper row is ships and the lower row is aircraft. (d) Schematic of the aircraft scattering point discretization. The red box is the ground truth and the green boxes are detection results. (e) Comparison of the same type of aircraft under different SAR imaging conditions.}   

	\label{discussion_comparision}
\end{figure*}

\begin{figure*}
	\centering
	\includegraphics[width=1.9\columnwidth]{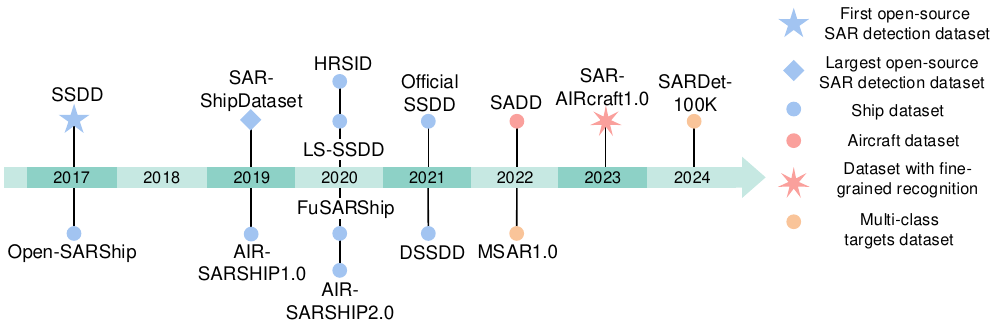}
	\caption{Milestone of SAR open source detection datasets. (SSDD \cite{li2017shipssdd}, Open-SARShip \cite{huang2017opensarship}, SAR-ShipDataset \cite{wang2019sarshipdataset}, AIR-SARSHIP1.0 \cite{xian2019airsarship}, HRSID \cite{wei2020hrsid}, 
 FuSARShip \cite{hou2020fusarship}, Official SSDD \cite{zhang2021sarssdd}, DSSDD \cite{hu2021dssddDual}, SADD \cite{zhang2022saddsefepnet}, SAR-AIRcraft1.0 \cite{zhirui2023sarcraft}, SARDet-100K \cite{li2024sardet100}.) \protect\footnotemark[5]}
	\label{sar_dataset_milestone}
\end{figure*}

\footnotetext[5]{LS-SSDD : \url{https://radars.ac.cn/web/data/getData?newsColumnId=6b535674-3ce6-43cc-a725-9723d9c7492c}, \\
AIR-SARSHIP2.0 : \url{https://radars.ac.cn/web/data/getData?newsColumnId=1e6ecbcc-266d-432c-9c8a-0b9a922b5e85}, \\
MSAR1.0 : \url{https://radars.ac.cn/web/data/getData?dataType=MSAR}.}

\subsection{Discussion}
According to experimental results, the overall detection performance of the model on the SAR ship target is better than that of the SAR aircraft target. Analyzing the performance differences between ship and aircraft targets in SAR image detection can help us understand the characteristics of datasets more deeply, optimize resource allocation, and ultimately improve the accuracy and efficiency of SAR image interpretation.

We analyze and summarize the reasons for the big difference between aircraft and ship target detection performance from four aspects: imaging background, target characteristics, dataset composition, and dataset size. We hope to provide some insights and enlightenments for the future development of the SAR special model and breakthrough of detection performance bottleneck.

\textbf{Imaging background.} As illustrated in Fig. \ref{discussion_comparision} (a), ships generally exist in both expansive maritime regions and coastal areas. Because ships have much higher grayscale properties than the surrounding sea clutter, their overall bright spot can be seen more clearly in open sea environments. In contrast, aircraft targets are usually found at airports, where the background complexity is significantly higher than that of ships. Strong scattering from man-made buildings like hangars and terminals makes it harder to distinguish aircraft from the surrounding infrastructure.

\textbf{Target characteristics.} The scattering points of ships are more concentrated and ships moving across the surface of the sea create wakes that reveal important target information. In contrast, the complex structure and scattering mechanisms associated with aircraft result in discrete and incoherent features, presenting several isolated scattering points, as illustrated in Fig. \ref{discussion_comparision} (a), (b), and (c), where the ellipse circled in the figure denotes the wake of the ship. Moreover, as Fig. \ref{discussion_comparision} (d) illustrates, aircraft edge contours are frequently ill-defined and their details lacking, making complete target detection more difficult. Additionally, aircraft are more sensitive to azimuth angles because of their complex structures, increasing the difficulty of accurate detection. Examples shown in Fig. \ref{discussion_comparision} (e) depict identical aerial refueling aircraft KC-135: with HISEA-1 imaging presented in the first row and GaoFen-3 imaging displayed below it.

\textbf{Dataset composition.} The ship datasets typically comprise a single class of ships, with no requirement for fine-grained recognition. This can be conceptualized as a binary classification task, where the primary objective is to differentiate the target from the background, thereby simplifying the detection process. In contrast, the aircraft detection dataset encompasses fine-grained recognition tasks associated with identified targets, which significantly increases the complexity of detection.

\textbf{Dataset size.} Currently, the majority of open-source SAR detection datasets pertain to ships, with relatively limited availability of aircraft detection datasets. Fig. \ref{sar_dataset_milestone} shows the current open-source SAR detection datasets. For aircraft detection, obtaining large-scale, high-quality labeled SAR image datasets presents considerable difficulties and manual labor. Furthermore, the absence of a comprehensive SAR target benchmark dataset hinders the effective training and evaluation of algorithms, thereby imposing limitations on research in the domain of target detection and recognition to some extent.

\section{Conslusion}\label{conclusions}

In this paper, we propose a novel MaDiNet for accurate target detection in SAR images, utilizing a diffusion model with Mamba. To accommodate the sparsity of SAR images, we introduce the diffusion model and frame SAR target detection as the task of generating the spatial position (center coordinates) and sizes (width and height) of target bounding boxes. To address challenges arising from discrete features of SAR targets and complex backgrounds, we have developed the mambaSAR module based on state-space models to capture comprehensive spatial structural information of targets. Extensive experiments conducted on three representative datasets validate the effectiveness of our method, demonstrating superior detection accuracy compared to existing methods. In the field of SAR detection, ship and aircraft targets are just a small part, and target detection still presents challenges. Therefore, future work will concentrate on delving deeper into the interconnection between diffusion models and SAR image characteristics, while expanding to multi-class targets, which is crucial for advancing the intelligent interpretation of SAR images.


\bibliographystyle{elsarticle-num}
\bibliography{cas-refs}

\end{document}